\newif\iffull
\newif\iffsen
\fulltrue
\fsenfalse
\documentclass[fleqn]{article}
\usepackage[a4paper]{geometry}

\usepackage[english]{babel}
\usepackage[T1]{fontenc}
\usepackage[utf8]{inputenc}
\usepackage{csquotes}
\usepackage{amsmath}
\usepackage{environ}
\usepackage{amssymb}
\usepackage{breqn}
\usepackage{mathtools}
\usepackage{stmaryrd}
\usepackage{hyperref}
\usepackage{amsthm}
\usepackage{bbm}
\usepackage{thmtools,thm-restate} 
\usepackage[capitalise,noabbrev]{cleveref}
\usepackage{etoolbox}
\usepackage{graphicx}
\usepackage[indLines=false]{algpseudocodex}
\usepackage{algorithm}
\usepackage{subcaption}
\usepackage{tikz}
\usetikzlibrary{arrows.meta, positioning}
\usepackage{xparse}
\usepackage{aligned-overset}
\usepackage{booktabs}
\usepackage{lscape}
\usepackage{longtable}
\usepackage{pifont} 

\NewCommandCopy{\proofqedsymbol}{\qedsymbol}
\AtBeginEnvironment{proof}{\renewcommand{\qedsymbol}{\ensuremath{\color{mDarkGray}\blacksquare}}}
\AtBeginEnvironment{example}{\renewcommand{\qedsymbol}{\ensuremath{\color{mDarkGray}\blacksquare}}}
\newlength{\savedparindent}
\theoremstyle{plain}
\newtheorem{theorem}{Theorem}[section]

\newtheorem{lemma}[theorem]{Lemma}
\newtheorem{corollary}[theorem]{Corollary}
\newtheorem{proposition}[theorem]{Proposition}
\newtheorem{remark}[theorem]{Remark}

\makeatletter
\let\oldendproof\endproof
\def\endproof{\oldendproof\aftergroup\@afterindentfalse\aftergroup\@afterheading}
\let\oldendtheorem\endtheorem
\def\endtheorem{\oldendtheorem\aftergroup\@afterindentfalse\aftergroup\@afterheading}
\let\oldendlemma\endlemma
\def\endlemma{\oldendlemma\aftergroup\@afterindentfalse\aftergroup\@afterheading}
\let\oldendcorollary\endcorollary
\def\endcorollary{\oldendcorollary\aftergroup\@afterindentfalse\aftergroup\@afterheading}
\makeatother

\theoremstyle{definition}
\newtheorem{definition}[theorem]{Definition}
\Crefname{definition}{Definition}{Definitions}
\makeatletter
\let\oldenddefinition\enddefinition
\def\enddefinition{\oldenddefinition\aftergroup\@afterindentfalse\aftergroup\@afterheading}
\makeatother

\theoremstyle{remark}
\newtheorem{examplex}[theorem]{Example}
\newenvironment{example}
{\pushQED{\qed}\renewcommand{\qedsymbol}{\rotatebox[origin=c]{90}{$\color{mDarkGray}\blacktriangle$}}\examplex}
{\popQED\endexamplex}
\makeatletter
\let\oldendexample\endexample
\def\endexample{\oldendexample\aftergroup\@afterindentfalse\aftergroup\@afterheading}
\makeatother

\Crefname{example}{Example}{Examples}
\Crefname{lstlisting}{Listing}{Listings}

\newcommand{\JBNote}[2]{\todo[size=tiny,color=mPurple,#2]{JB: #1}}

\NewDocumentEnvironment{joreomit}{ +b }{}{}

\newcommand{\mon}{\tau}
\newcommand{\dbrack}[1]{\llbracket{} #1 \rrbracket{}}
\newcommand\defeq{\mathrel{\overset{\makebox[0pt]{\mbox{\normalfont\tiny\sffamily def}}}{=}}}
\newcommand{\grameq}{::=}
\newcommand{\seteq}{\coloneqq}
\newcommand{\true}{\mathsf{true}}
\newcommand{\false}{\mathsf{false}}
\newcommand{\pvi}{\mathtt{PVI}}
\newcommand{\pv}[2]{\pvi_{#2}(#1)}
\newcommand{\data}[2]{\mathtt{arg}_{#2}(#1)}
\newcommand{\fv}{\ensuremath{\mathtt{free}}}

\newcommand{\pred}[2]{\mathtt{name}_{#2}(#1)}
\newcommand{\pviPos}{N}

\newcommand{\pviPositions}[3]{\pviPos_{#1(#2)}(#3)}
\newcommand{\pviPositionsA}[2]{\pviPos_{#1}(#2)}
\newcommand{\simultaneousChainedG}{G_{X,i}}
\newcommand{\npred}{\#\pvi}
\newcommand{\npredset}{N_\pvi}
\newcommand{\occ}{\mathtt{occ}}
\newcommand{\bnd}{\mathtt{bnd}}
\newcommand{\sgn}{\mathtt{sgn}}

\newcommand{\fp}[2]{{#1}_{\langle #2 \rangle}}
\newcommand{\pvisub}[2]{[#1 / #2]}
\newcommand{\exprsub}[2]{[#1 \seteq #2]}
\newcommand{\isub}[2]{[#1 \mapsto #2]}
\newcommand{\fvsub}[1]{\langle #1 \rangle}
\newcommand{\isubsim}[3]{[#3 \mapsto #2]}
\newcommand{\lc}[1]{\{\, #1 \,\}}
\newcommand{\lb}[1]{(\, #1 \,)}
\newcommand{\unfold}[2]{\mathtt{chain}_{#2}(#1)}
\newcommand{\unfoldAll}[2]{\mathtt{chain}_{#2}(#1)}

\newcommand{\reason}[1]{\text{[#1]}}
\newcommand{\tool}[1]{\lstinline|#1|}
\newcommand{\pbeschain}{\tool{pbeschain}}
\newcommand{\overs}[2]{\stackrel{\ensuremath{#1}}{#2}}
\newcommand{\orand}{\diamond}
\newcommand{\sort}{{:}}
\newcommand{\tland}{\text{ and }}
\newcommand{\tlor}{\text{ or }}
\newcommand{\timplies}{\text{ implies }}
\newcommand{\tequals}{\text{ equals }}
\newcommand{\tforall}[1]{\text{for all }#1,}
\newcommand{\texists}[1]{\text{for some }#1,}
\renewcommand{\vec}[1]{\mathbf{#1}}

\newcommand{\ie}{\emph{i.e.}}
\newcommand{\cf}{\emph{cf.}}
\newcommand{\quantifier}{\mathcal{Q}}

\usepackage{todonotes}
\setuptodonotes{color=mOrange,textcolor=mLightBg}

\NewDocumentEnvironment{myomit}{ +b }{}{}

\title{Selective Chaining of \\
Parameterised Boolean Equation Systems}
\author{Jore J. Booy \and Tim~A.C.~Willemse}

\usepackage{xcolor}
\definecolor{mGreen}{HTML}{8fbcbb}
\definecolor{mDarkGray}{HTML}{4c566a}
\definecolor{mDarkestWhite}{HTML}{d8dee9}
\definecolor{mLightBg}{HTML}{E6EFEF}
\definecolor{mGray}{rgb}{0.5,0.5,0.5}
\definecolor{mPurple}{rgb}{0.58,0,0.82}
\definecolor{mBlue}{HTML}{5e81ac}
\definecolor{mOrange}{HTML}{d08770}
\definecolor{mYellow}{HTML}{ebcb8b}
\definecolor{mRed}{HTML}{bf616a}

\hypersetup{
    colorlinks=true,
    linkcolor=mBlue,
    filecolor=magenta,
    urlcolor=mBlue,
    citecolor=mBlue,
  }

\let\oldforall\forall
\renewcommand{\forall}{\oldforall \, }

\let\oldexist\exists
\renewcommand{\exists}{\oldexist \: }

\usepackage{listings}

\lstdefinestyle{mcrl2}{
  language=,
  basicstyle=\ttfamily\small\lst@ifdisplaystyle\footnotesize\fi,
  keywords={sort,act,proc,init,sum,allow,hide,comm,block,rename,var,eqn,map,struct,cons,forall,exists,glob,pbes},
  morekeywords={Set,Nat,Bool,List,true,false,whr,end,delta,tau,mu,nu},
  keywordstyle=\bfseries\color{mBlue},
  stringstyle=\color{mDarkGray},
  commentstyle=\color{mOrange},
  sensitive=true,
  morecomment=[l]{\%},
  captionpos=b,
  commentstyle=\itshape,
  stringstyle=\ttfamily,
  showstringspaces=false,
  numbers=none,
  numberstyle=\tiny,
  breaklines=true,
  frame=single,
  mathescape=true
}

\usepackage[
  backend=biber,
  style=alphabetic,
  sorting=ynt
]{biblatex}
\usepackage{listings}

\begin{document}

\maketitle

\begin{abstract}
\noindent Parameterised Boolean Equations Systems (PBESs) are used to verify properties of formal industrial models. 
We introduce a technique called \emph{selective chaining} to simplify these PBESs by unfolding specific predicate variable instances when the unfolding results in a PBES that is likely to be simpler.
For determining the latter we propose a set of heuristics.
We show the formal correctness of the chaining operator and provide theoretical upper bounds using relevancy graphs on the size of the verification problem after chaining. 
Lastly, we study the effectiveness of our heuristics, showing that selective chaining can reduce memory usage and yield significant speed-ups solving PBESs that encode model checking problems for complex industrial systems.
\end{abstract}

\section{Introduction}
Establishing the correctness of software embedded in critical devices and controlling systems is typically a necessity, but it is hard to get formal guarantees.
Tools such as TLA$^+$~\cite{lamport2002specifying} and mCRL2~\cite{bunteMCRL2ToolsetAnalysing2019b} allow one to model the design of such software, and subsequently verify its properties using a technique called \emph{model checking}.
This helps to exclude critical design flaws in the early stages of the design, when fixing issues is still relatively cheap.

Model checking works by inspecting all possible configurations that the software can be in, and analysing how these affect the properties one expects the system to uphold.
However, a classical problem with model checking is the infamous \emph{state explosion problem}, which manifests itself by the analysis technique requiring considerable computational effort and memory space, limiting the applicability of the analysis technique.
Methods that help to (partly) mitigate the state explosion problem are therefore much in demand.

One uniform way to solve the model checking problem is through the use of Parameterised Boolean Equation Systems~\cite{grooteParameterisedBooleanEquation2005}.
These are sequences of first-order fixpoint equations of the shape $\sigma X(d_1\sort D_1,\dots,d_n\sort D_n) = \phi$, where $\sigma$ is either a least or greatest fixpoint, $X$ is a predicate variable, $d_i$ are data parameters of type $D_i$ and $\phi$ is a first-order formula; for example,  $\lb{ \nu X(n\sort N) = \forall m\sort N. ((n > m) \vee X(n+1)) }$ is an equation system consisting of a single equation.

A~sound transformation on PBESs is to unfold its equations: the \emph{rolling rule} of fixpoint theory, sometimes stated as $\sigma (f \circ g) = f(\sigma (g \circ f))$, see also~\cite{neeleOperationsFixpointEquation2024a}, dictates that one can unwind the right-hand side of an equation $\sigma X(d_1\sort D_1,\dots,d_n\sort D_n)= \phi$ by replacing all occurrences of $X$ in $\phi$ by $\phi$ itself.
Such unfolding sometimes helps to reduce the computational effort needed to solve the PBES, and thereby the model checking problem, but it can also have quite the opposite effect, as demonstrated by the following equation:
\[\lb{ \nu X(n\sort N) = ((n > 4) \vee X(n+1)) \wedge X(n+2) }\]
When unfolding this equation, the subformula $X(n+1)$ will be replaced by the right-hand side of the equation with parameter $n$ replaced by $n+1$, so $(n+1 > 4 \vee X(n+1+1)) \wedge X(n+1+2)$. Similarly, subformula $X(n+2)$ will be replaced by $(n+2 > 4 \vee X(n+2+1) ) \wedge X(n+2+2)$.
After minor simplification, unfolding will thus result
\iffsen
\else
in the following equation:
\[\lb{ \nu X(n \sort N) = ( (n > 4) \vee ( (n > 3 \vee X(n+2)) \wedge X(n+3) ) ) \wedge ( (n > 2 \vee X(n+3)) \wedge X(n+4)) }\]
\fi

In this paper, we introduce a refinement of the unfolding operation that we call \emph{selective chaining}.
It entails unfolding only specific predicate variable instances to simplify the PBES, while leaving others unaffected.
Our first contribution is a proof of soundness of selective chaining.
The soundness proof reveals an upper bound on the increase in size of the certificate of a solution to a PBES after chaining.
Moreover, we show that the search space that contains this certificate exhibits a similar maximal size increase, proving that adverse effects of selective chaining, should they manifest, are limited.
Our second contribution is a set of heuristics that allow for tuning the degree of chaining, thereby largely sidestepping the adverse effects of unfolding altogether.
A third, and final, contribution is an assessment of the effectiveness of selective chaining, and our heuristics, for solving real-world model checking problems.
For this, we use a set of experiments that consists of models for two pieces of complex control software for infrastructure, and software running a component used in factory automation systems, and properties that capture highly non-trivial safety and liveness requirements.
Our experiments reveal that PBESs can be solved up-to 25$\times$ faster, and require up-to 14$\times$ less memory.



\paragraph{Related work.} Various methods of simplifying PBESs have been proposed over the years. Static methods include removing redundant parameters and constant parameters~\cite{orzanStaticAnalysisTechniques2009}, more advanced liveness analysis of parameters~\cite{keirenImprovedStaticAnalysis2014}, and manipulating quantifiers to reduce computational effort~\cite{neeleRemovingQuantifiersSimplify2022}.
Our effort is complementary to these techniques by operating  on predicate variable instances instead.
In~\cite{kobayashiFoldUnfoldTransformations2020}, fold and unfold transformations are defined for the MuArith fixpoint logic.
The unfolding operation is exactly the rolling rule described earlier, and is similar to the substitution described in~Lemma~18 of~\cite{grooteParameterisedBooleanEquation2005}, replacing all occurrences of a predicate with a specific name with its right-hand side. We instead consider unfolding any single PVI or set of PVIs with the same name and arguments.

\paragraph{Outline.}
We cover the preliminaries in \cref{sec:preliminaries}. In \cref{sec:chaining}, we define chaining, prove it preserves the solution of the original PBES and gives theoretical upper bounds on the size of the resulting state space. In \cref{sec:heuristics}, we cover the different heuristics used to allow or disallow chaining, and describe the algorithm that applies selective chaining to an arbitrary PBES. In \cref{sec:practical_effectiveness}, we evaluate the effectiveness of the heuristics using real-world case studies.

\section{Preliminaries}
\label{sec:preliminaries}
We assume a set of sorts $D, D', D_1, D_2,\ldots$, representing domains $\mathbb{D}, \mathbb{D}', \mathbb{D}_1, \mathbb{D}_2, \ldots$
We write $B$ to denote the sort of Booleans, and assume that $B$ has two distinct elements $\top$ and $\bot$, representing values $\true$ and $\false$.
The sort of natural numbers is denoted by $N$.
We assume all the standard operators defined on $B$ and $N$, including but not limited to $\neg$, $\approx$, and $\leq$.
Furthermore, we assume an infinite set $\mathcal{D}$ of sorted data variables; given a variable $d \in \mathcal{D}$, we write $d \sort D$ to indicate that $d$ is of sort $D$.
We use $\vec{d}$ to denote a list of variables, $\vec{D}$ to denote the sorts of these, and $\mathbbm{D}$ to denote the domain of these.
A \emph{data environment} is a mapping $\delta$ that assigns a value from the domain $\mathbb{D}$ to each variable $d\sort D$.
Data environment $\delta[v/d]$ assigns value $v$ to variable $d$ and $\delta(d')$ to variables $d' \neq d$.
Data terms are typically denoted by $e, e_1, e_2, \ldots$, and we write $\fv(e)$ to denote the set of free data variables occurring in $e$.
We assume an interpretation function that, together with a data environment $\delta$ assigning values to the data variables, maps data terms $e$ to the value they represent; we write $\dbrack{e}\delta$ to denote the interpretation of $e$ in the context of $\delta$.
\iffsen
\else
\begin{definition}
  \label{def:pred-form}
  \emph{Predicate formulae} are defined through the following grammar:
  \[
    \varphi, \psi \grameq b \mid X(e_1,\ldots,e_n) \mid \varphi \land \psi \mid \varphi \lor \psi \mid  \forall d\sort D.\varphi \mid \exists d\sort D.\varphi
  \]
  in which $b$ is a term of sort $B$, $X$ of sort $D_1 \times \cdots \times D_n \to B$ is a predicate variable taken from an infinite set of sorted predicate variables $\mathcal{P}$, $e_i$ are data terms of sort $D_i$,
  and $d$ is a data variable of sort $D$.
\end{definition}
\fi
We generalize function $\fv$ to predicate formulae in the expected way; we write $\fp{\varphi}{d_1,\ldots,d_n}$ to indicate that $\varphi$ is a predicate formula satisfying $\fv(\varphi) \subseteq \{d_1,\ldots,d_n\}$.
Let $\varphi\exprsub{(d_1,\ldots,d_n)}{(e_1,\ldots,e_n)}$ denote the simultaneous syntactic substitution of terms $e_i$ for variables $d_i$ in $\varphi$. We shorten $\fp{\varphi}{d_1,\ldots,d_n}\exprsub{(d_1,\ldots,d_n)}{(e_1,\ldots,e_n)}$ to $\varphi\fvsub{e_1,\ldots,e_n}$.
A predicate formula $\fp{\varphi}{d_1,\ldots,d_n}$ is \emph{well-named} iff all variables in quantifier definitions $(d',d'',\ldots)$ in $\varphi$ are pairwise distinct and disjoint from $(d_1,\ldots,d_n)$. We consider only well-named predicate formulae.
A (sub)formula of the shape $X(e_1,\ldots,e_n)$ is referred to as a \emph{predicate variable instance} (PVI), and we call the predicate variable $X$ the \emph{name} of $X(e_1,\ldots,e_n)$.
We use $\vec{d}_X$ to denote the list of parameters of $X$ with sort $\vec{D}_X$.
Let $\occ(\varphi)$ denote the set of predicate variables that occur in predicate formula $\varphi$, and $\pvi(\varphi)$ denote the set of PVIs in $\varphi$.
A predicate formula is \emph{simple} iff it contains no predicate variable instances.
We use $f \implies \varphi$ as a shorthand for $\neg f \lor \varphi$, where $\neg$ is part of the data term.
A predicate formula is a \emph{standard recursive clause} iff it has the shape:
\[
  \begin{split}
    \bigvee_{j \in J_{X_i}} \exists e_j \sort E_j. f_j(\vec{d}, e_j) \land X_j(\vec{g}_j(\vec{d}, e_j)) ~~~\text{ or }~~~
    \bigwedge_{j \in J_{X_i}} \forall e_j \sort E_j. f_j(\vec{d}, e_j) \implies X_j(\vec{g}_j(\vec{d}, e_j))
  \end{split}
\]

\iffull
\begin{definition}
  \label{def:inter-predicate}
  \fi
  Let $\eta$ be a predicate environment, assigning a set $V \subseteq \mathbbm{D}_X$ to predicate variables $X \sort \vec{D}_X \to B$.
  The \emph{interpretation} of a predicate formula $\varphi$ in the context of predicate environment $\eta$ and data environment $\delta$ is denoted as $\dbrack{\varphi}\eta\delta$.
  It is defined inductively as follows:
  \[\arraycolsep=2.5pt 
    \begin{array}{lclclcl}
      \dbrack{b}\eta\delta &\defeq& \dbrack{b}\delta & &  \dbrack{X(e_1,\ldots,e_n)}\eta\delta &\defeq & \langle\dbrack{e_1}\delta, \ldots, \dbrack{e_n}\delta\rangle \in \eta(X) \\
      \dbrack{\varphi \land \psi}\eta\delta &\defeq & \dbrack{\varphi}\eta\delta \tland \dbrack{\psi}\eta\delta & &
      \dbrack{\forall d\sort D.\varphi}\eta\delta &\defeq & \tforall{v\in\mathbb{D}} \dbrack{\varphi}\eta\delta\lbrack v / d \rbrack \\
      \dbrack{\varphi \lor \psi}\eta\delta &\defeq &   \dbrack{\varphi}\eta\delta \tlor \dbrack{\psi}\eta\delta & &
      \dbrack{\exists d\sort D.\varphi}\eta\delta &\defeq & \texists{v\in\mathbb{D}} \dbrack{\varphi}\eta\delta\lbrack v / d \rbrack \\
  \end{array}\]
  \iffull
\end{definition}
\fi

Given predicate environments $\eta, \eta'$ we write $\eta \subseteq \eta'$ iff for all $Y \in \mathcal{P}$ we have $\eta(Y) \subseteq \eta'(Y)$.
We write $\eta\pvisub{b}{X(v)}$ with $b \in B, X \in \mathcal{P},v\in \mathbbm{D}_X$ to denote an updated predicate environment:
\[
  \eta\pvisub{b}{X(v)}(Y) =
  \begin{cases}
    \eta(Y) & \text{if } Y \neq X \\
    \eta(Y)\cup \{w\}& \text{if } Y = X\text{ and } b = \top \\
    \eta(Y)\setminus \{w\} & \text{if } Y = X\text{ and } b = \bot
  \end{cases}
\]
Parameterised Boolean Equation Systems~\cite{grooteParameterisedBooleanEquation2005}, or equation systems for short, are sequences of fixpoint equations ranging over predicate formulae.
\iffull
\begin{definition}
  \fi
  \emph{Equation systems} are defined through the following grammar:
  \[
    \mathcal{E} \grameq \emptyset
    \mid
    \lb{\sigma X(d_1\sort D_1, \ldots, d_n \sort D_n) = \fp{\varphi}{d_1,\ldots,d_n}}\, \mathcal{E}
  \]
  where $\emptyset$ denotes the empty sequence, $\varphi$ is a predicate formula, $\sigma \in \lc{\mu,\nu}$ is a least/greatest fixpoint sign, and~$X \in \mathcal{P}$.
  An equation system is in \emph{standard recursive form} (SRF) when all right-hand sides are standard recursive clauses.
  \iffull
\end{definition}
\fi
A trailing $\emptyset$ is typically omitted for non-empty equation systems.
Henceforth, we write $\sigma$ to denote either the least fixpoint $\mu$ or the greatest fixpoint $\nu$.
For a given equation system~$\mathcal{E}$, we typically use $\sigma_X$ to denote the fixpoint symbol for the equation for $X$ in $\mathcal{E}$, and $\varphi_X$ to denote the predicate formula occurring in the right-hand side of the equation for $X$.
We write $\occ(\mathcal{E})$ to denote the union of all predicate variables occurring in the right-hand side of the equations in~$\mathcal{E}$, generalizing $\occ(\varphi)$.
The set of \emph{binding predicate variables}, denoted $\bnd(\mathcal{E})$, is the set of predicate variables occurring at the left-hand side of the equations in~$\mathcal{E}$.
An equation system~$\mathcal{E}$ is said to be \emph{well-formed} when every binding predicate variable occurs in exactly one equation's left-hand side in~$\mathcal{E}$, and it is \emph{closed} when $\occ(\mathcal{E}) \subseteq \bnd(\mathcal{E})$. 
We only consider well-formed equations systems.

The solution of an equation system~$\mathcal{E}$ can be characterized by means of a denotational semantics, which we forego defining. Instead, we define the solution of equation systems in terms of \emph{proof graphs}~\cite{cranenProofGraphsParameterised2013}: a graph structure that explains the solution to a set of predicate variable instances in terms of their dependencies on other predicate variable instances and that provide an intuition for the solution of the equation system.
For $X$ with arguments of sort $\vec{D}_X$, we denote the \emph{signature} of $X$ by $\sgn(X)$, and define it as the set of all elements taken from $\mathbbm{D}_X$.
The signature of an equation system~$\mathcal{E}$, denoted $\sgn(\mathcal{E})$, is defined as $\lc{ (X, \vec{v})\mid  X \in \bnd(\mathcal{E}), \vec{v} \in \sgn(X) }$.
A~set $S \subseteq \sgn(\mathcal{E})$ induces environments $S_{\true}$ and $S_{\false}$ as follows: for all $X \in \mathcal{P}$,
$S_{\true}(X) = \lc{\vec{v}  \mid (X,\vec{v}) \in S }$ and
$S_{\false}(X) = \lc{\vec{v}  \mid (X,\vec{v}) \in \sgn(\mathcal{E})\setminus S }$.
These environments are used to define two types of dependency graphs:
depending on the polarity ($\true$ or $\false$), such a graph captures a consistent truth assignment to a set of predicate variable instances such that these align with the evaluation of their associated right-hand side formulas.
Formally, a true-dependency graph and a false-dependency graph are defined as follows.

\begin{definition}
  \iffsen
\else
  Let~$\mathcal{E}$ be a closed equation system.
  A directed graph $(V,E)$, with $V \subseteq \sgn(\mathcal{E})$, is a:
  \begin{itemize}
    \item \emph{true-dependency graph} if for all $(X,\vec{v}) \in V$ and for all data environments $\delta$ it holds that $\dbrack{\varphi_X} ( (X,\vec{v})^\bullet)_{\true} \delta[\vec{v}/\vec{d}_X] = \true$, and
    \item \emph{false-dependency graph} if for all $(X,\vec{v}) \in V$ and for all data environments $\delta$ it holds that $\dbrack{\varphi_X}( (X,\vec{v})^\bullet)_{\false}\delta[\vec{v}/\vec{d}_X] = \false$.
  \end{itemize}
  where $(X,\vec{v})^\bullet = \lc{ (Y,\vec{w}) \mid \lb{ (X,\vec{v}), (Y, \vec{w}) } \in E }$ is the set of immediate $E$-successors of $(X,\vec{v})$.
  \fi
\end{definition}
Let $\triangleleft_{\mathcal{E}}$ be the total order on the binding predicate variables in~$\mathcal{E}$, obtained by the order in which their respective equations occur in~$\mathcal{E}$.
For instance, we have $X \triangleleft_{\mathcal{E}} Y$ for $\mathcal{E} = \lb{ \nu X(n\sort N) = Y(n+1)}\lb{ \mu Y(k\sort N) = X(k) \wedge Y(k+1) }$.

\begin{definition}\label{def:proof_graphs}
  \iffsen
\else
  A true-dependency graph $G = (V,E)$ for~$\mathcal{E}$ is a \emph{proof graph} iff for all infinite paths $\pi \in V^\omega$ in $G$, the equation associated with the smallest (w.r.t.\ $\triangleleft_{\mathcal{E}}$) predicate variable occurring infinitely often on $\pi$ is a $\nu$-equation.
  Likewise, a false-dependency graph $G = (V,E)$ for~$\mathcal{E}$ is a \emph{refutation graph} iff for all infinite paths $\pi \in V^\omega$ in $G$, the equation associated with smallest (w.r.t.\ $\triangleleft_{\mathcal{E}}$) predicate variable occurring infinitely often on $\pi$ is a $\mu$-equation.
  \fi
\end{definition}
The following definition gives the semantics of a PBES in terms of proof graphs.

\begin{definition}
  \label{def:proof-graph-means-true}
  Let~$\mathcal{E}$ be a closed equation system, and let $X \in \bnd(\mathcal{E})$.
  Then $X(\vec{v})$ holds in~$\mathcal{E}$  iff there is a proof graph $(V,E)$ such that $(X,\vec{v}) \in V$; and $X(\vec{v})$ does not hold in~$\mathcal{E}$  iff there is a refutation graph $(V,E)$ such that $(X,\vec{v}) \in V$.
  The proof graph or refutation graph showing that $X(\vec{v})$ holds true or not is called a \emph{certificate}.
\end{definition}



\section{Chaining}
\label{sec:chaining}

Computing the solution to a specific predicate variable instance can be done in various ways, but a typical way of doing so involves computing the set of predicate variable instances potentially relevant for its solution.
Determining this set can be done in a way that is akin to an explicit graph exploration.
For instance, determining the solution to $X(0)$ in the equation system $\lb{\nu X(n\sort N) = (n > 4) \vee X(n+1)}$ can be done by establishing that $X(0)$ depends on $X(1)$; in turn, determining the solution to $X(1)$ requires establishing that $X(1)$ depends on $X(2)$; and so on until one determines that $X(5)$ is simply $\true$ because $5 > 4$.
A \emph{relevancy graph}~\cite{liemExtractionInvariantsParameterised2023}, which we recall below, provides a proxy for the size of this space.

\begin{definition}
  \iffsen
\else
  Let~$\mathcal{E}$ be a closed equation system. A directed graph $(V,E)$, with $V \subseteq \sgn(\mathcal{E})$, is a \emph{relevancy graph} if for all $(X,\vec{v}) \in V$ and $(Y, \vec{w}) \in \sgn(\mathcal{E})$, we have $((X,\vec{v}), (Y,\vec{w})) \in E$ iff $\exists \eta,\delta : \dbrack{\varphi_X}\eta\pvisub{\top}{Y(\vec{w})}\delta[\vec{v}/\vec{d}_X] \neq \dbrack{\varphi_X}\eta\pvisub{\bot}{Y(\vec{w})}\delta[\vec{v}/\vec{d}_X]$. Given some $v \in \sgn(\mathcal{E})$, a relevancy graph $(V,E)$ is a \emph{relevancy graph for $v$} iff $v \in V$.
  \fi
\end{definition}
From a relevancy graph, one can always construct a minimal relevancy graph.
In general, this graph, which represents the relevant set of predicate variable instances and their inter-dependencies, is sufficient to extract a proof graph, \emph{e.g.}, using Zielonka's recursive algorithm~\cite{DBLP:journals/tcs/Zielonka98} for solving parity games.

Computing complicated chains of dependencies can be quite costly, as it involves evaluating complex right-hand side formulas and extracting the relevant predicate variable instances.
Moreover, some of this work is rather repetitive; in case of the given example, the process of determining the relevant predicate variable instances for when $n$ is $0$ is the same as for when $n$ is $1$ or $2$.
Rather than computing these dependencies for each such instance separately and on-the-fly in a graph exploration, we may accelerate these computations by observing that in general, one may syntactically unfold the right-hand side of an equation.
The mathematical underpinning of such an unfolding is given by the \emph{rolling rule} $\sigma (f \circ g) = f(\sigma (g \circ f))$ of fixpoint theory~\cite{neeleOperationsFixpointEquation2024a}, to which we already alluded in the introduction.
The rather blunt unwinding of this rule, however, often has the undesirable effect of actually increasing the complexity of the equation system rather than simplifying it, as the example in the introduction shows.
The technique we introduce next refines the unwinding and instead operates locally, with the aim to limit the blow-up as much as possible.

\subsection{The Chaining Operator}
\label{sec:chaining_operator}

Given a predicate formula $\varphi$, we write $\npred(\varphi)$ to indicate the number of predicate variable instances in $\varphi$ and $\npredset(\varphi)$ to indicate the set $\lc{i \mid 1 \leq i \leq \npred(\varphi)}$.
We write $\pv{\varphi}{i}$ to denote the predicate variable instance at position $i$ in $\varphi$.
Note that $\pv{\varphi}{i}$ is of the shape $X(e_1,\ldots,e_n)$ for some $X \in \mathcal{P}$ of arity $n$ and each of the data terms $e_1,\ldots,e_n$ are of the appropriate sort.
To facilitate notation elsewhere, we let $\pred{\varphi}{i}$ denote this predicate variable name $X$ and let $\data{\varphi}{i}$ denote the list of arguments $e_1, \ldots, e_n$, so $\pv{\varphi}{i} = \pred{\varphi}{i}(\data{\varphi}{i})$.
The following definition is a generalization of the local substitution operator $\varphi\isub{i}{\psi}$ of~\cite{keirenImprovedStaticAnalysis2014}.
\begin{definition}
  \iffsen
\else
  Let $\psi$ be a predicate formula and let $I\subseteq\npredset(\varphi)$. Then $\varphi\isub{I}{ \psi}$ is defined inductively as follows:
  \[
    \begin{split}
      b\isubsim{j}{\psi}{I} &\defeq b,\\
      Y(\vec e)\isubsim{j}{\psi}{I} &\defeq
      \begin{cases}
        \psi & \text{if }1\in I\\
        Y(\vec e) & \text{otherwise}
      \end{cases}\\
      (\varphi_1\land\varphi_2)\isubsim{j}{\psi}{I} &\defeq
      \varphi_1\isubsim{j}{\psi}{\lc{i\in I\mid i\leq\npred(\varphi_1)}}\\
      &\qquad \land
      \varphi_2\isubsim{j}{\psi}{\lc{i-\npred(\varphi_1)\mid i\in I\land i>\npred(\varphi_1)}}\\
      (\varphi_1\lor\varphi_2)\isubsim{j}{\psi}{I} &\defeq
      \varphi_1\isubsim{j}{\psi}{\lc{i\in I\mid i\leq\npred(\varphi_1)}}\\
      &\qquad \lor
      \varphi_2\isubsim{j}{\psi}{\lc{i-\npred(\varphi_1)\mid i\in I\land i>\npred(\varphi_1)}}\\
      (\forall d\sort D.\varphi)\isubsim{j}{\psi}{I} &\defeq
      \forall d\sort D.\varphi\isubsim{j}{\psi}{I}\\
      (\exists d\sort D.\varphi)\isubsim{j}{\psi}{I} &\defeq
      \exists d\sort D.\varphi\isubsim{j}{\psi}{I}
    \end{split}
  \]
  \fi
\end{definition}
The operator has an expected identity property.
\begin{restatable}{lemma}{simultaneousSubstitutionEmpty}
  \label{lemma:simultaneous-substitution-empty}
  Let $\varphi,\psi$ be predicate formulae. Then $\varphi\isubsim{j}{\psi}{\varnothing}=\varphi$.
\end{restatable}

\noindent\ignorespacesafterend
For a PVI $Y(\vec e)$ occurring in $\varphi$, define the set of all its positions by:
\iffull
\[
  \pviPositions{Y}{\vec e}{\varphi}
  =\lc{i\in\npredset(\varphi)\mid\pred{\varphi}{i}=Y\land\data{\varphi}{i}=\vec e}.
\]
\else
$
\pviPositions{Y}{\vec e}{\varphi}
=\lc{i\in\npredset(\varphi)\mid\pred{\varphi}{i}=Y\land\data{\varphi}{i}=\vec e}.
$
\fi
We can replace the PVI from such a set with any other PVI in that set. This is formalized next.
\begin{restatable}{lemma}{simultaneousMapstoidentity}
  \label{lemma:simultaneous-mapsto-identity}
  Let $\varphi$ be a predicate formula, $i\in\npredset(\varphi)$ and $I\subseteq\pviPositionsA{\pv{\varphi}{i}}{\varphi}$.
  Then $\varphi\isubsim{i}{\pv{\varphi}{i}}{I}=\varphi$.
\end{restatable}
Now, we give a definition of chaining using these definitions.

\begin{definition}
  \label{def:all-occurrence-chaining}
  Let $X\in\bnd(\mathcal{E})$, let $i\in\npredset(\varphi_X)$ such that $\pred{\varphi_X}{i}=X$, and let $I\subseteq\pviPositions{X}{\data{\varphi_X}{i}}{\varphi_X}$.
  \iffsen
\else
  \emph{Chaining} set $I$ replaces the selected occurrences by the same instantiated right-hand side:
  \[
    \unfoldAll{\varphi_X}{I}
    =\varphi_X\isubsim{j}{\varphi_X\fvsub{\data{\varphi_X}{i}}}{I}.
  \]
  \fi
  We write $\unfoldAll{\mathcal{E},X}{I}$ for the equation system in which the right-hand side of $X$ is replaced by $\unfoldAll{\varphi_X}{I}$.
\end{definition}

\begin{example}
  \label{example:less-dependency-after-chaining}
  Reconsider the predicate formula $\varphi$ given by $( (n > 4) \vee X(n+1)) \wedge X(n+2)$.
  \iffsen
\else
  Observe that $\npred(\varphi) = 2$, i.e., there are exactly two predicate variable instances in $\varphi$.
  Applying chaining to the second of these leads to the formula $(n > 4 \vee X(n+1)) \wedge ((n > 2 \vee X(n+3)) \wedge X(n+4))$.
  \fi
  This simplifies the dependency chains of all predicate variable instances solely to trivial instances while limiting blow-up.
\end{example}

\subsection{Correctness of Chaining and Estimating the Solution Space}
\label{sec:chaining_operator_correctness}

\newcommand{\chainedG}{G_{X,i}}
\newcommand{\regularE}{E_{G}}
\newcommand{\tilE}[1][i]{\widetilde E_{G,X,#1}}
We establish the correctness of chaining by providing a proof graph for a chained PBES, given a proof graph for the original PBES. Analogous arguments can be used to show that if a refutation graph for the original PBES exists, it also exists for the chained PBES. From these observations, it then follows that the solution of the PBES is preserved after chaining.

We fix an equation system~$\mathcal{E}$ for the remainder of this section.
For the purpose of establishing the correctness of the transformation and estimating the potential size increase of the relevancy graphs induced by the transformation, we extend a given proof or relevancy graph $G$ with a set of edges~$\tilE$.
This set contains additional edges that might be needed to show that the extended graph~$\chainedG$ is a proof graph or relevancy graph of the transformed PBES.

\newcommand{\values}{\iffsen
\else
  \mathtt{succvals}
\fi}
\begin{definition}
  \label{def:values-egxi}
  Let $G = (V,\regularE)$ be a directed graph with $V \subseteq \sgn(\mathcal{E})$, let $X \in \bnd(\mathcal{E})$, and let $(X,\vec{v}) \in V$. Let $i \in \npredset(\varphi_X)$ be such that $\pred{\varphi_X}{i}=X$.
  The set of possible values associated with the arguments of the $i$-th predicate variable instance in~$\varphi_X$, denoted by $\values((X,\vec{v}),i,\regularE)$, is defined as:
  \[
    \iffull
    \values((X,\vec{v}),i,\regularE)=
    \fi
    \lc{\vec{u}\mid\exists\delta:
      \vec{u}=\dbrack{\data{\varphi_X}{i}}\delta[\vec{v}/\vec{d}_X]
    \land ((X,\vec{v}),(X,\vec{u}))\in\regularE}.
  \] 
  We define the \emph{extended graph} $\chainedG$ of $G$ as $\chainedG = (V,\regularE\cup\tilE)$,
  \iffsen
\else
  where
  \[
    \tilE=\lc{((X,\vec{v}),(Y,\vec{w}))\mid (X,\vec{v})\in V
      \land\exists\vec{u}\in\values((X,\vec{v}),i,\regularE):
    ((X,\vec{u}),(Y,\vec{w}))\in\regularE}.
  \]
  \fi
\end{definition}
The following two lemmata are used in our proof of~\cref{proposition:simultaneous-chaining-true-dependency-graph}.

\begin{restatable}{lemma}{fromdeltamovefreevarin}
  \label{lemma:from-delta-move-free-var-in}
  Let $\fp{\varphi}{\vec{d}}$ be a predicate formula with $\vec{d}\sort\vec{D}$,
  and let $\vec{e}$~be a data term of sort~$\vec{D}$.
  Then for all environments $\eta,\delta,\delta'$, $\dbrack{\varphi}\eta\delta'
  [\dbrack{\vec{e}}\delta/\vec{d}]
  = \dbrack{\varphi\fvsub{\vec{e}}}\eta\delta$.
\end{restatable}
\begin{restatable}{lemma}{simultaneousSubstitutionMonotonicity}
  \label{lemma:simultaneous-substitution-monotonicity}
  Let $\psi,\psi'$ be predicate formulae and $\eta,\eta'$ be predicate environments such that $\eta\subseteq\eta'$.
  Assume for all $\delta$ that $\dbrack{\psi}\eta\delta\timplies\dbrack{\psi'}\eta'\delta$.
  Then for all $\delta,\varphi$ and $I\subseteq\npredset(\varphi)$, we have
  $\dbrack{\varphi\isubsim{j}{\psi}{I}}\eta\delta\timplies\dbrack{\varphi\isubsim{j}{\psi'}{I}}\eta'\delta$.
\end{restatable}
We split the proof of correctness of chaining into two parts. First, we show the graph extended with the edges induced by the selected occurrences is a valid true-dependency graph, and then use this result to show it is a proof graph by reasoning that all infinite paths satisfy the proof graph condition.

\newif\ifapxpropproofgraph
\iffsen
\else
\apxpropproofgraphfalse
\fi
\newif\ifapxthmproofgraph
\iffsen
\else
\apxthmproofgraphfalse
\fi

\newcommand{\depgraphX}{Y}
\begin{restatable}{proposition}{simultaneousChainingTrueDependencyGraph}
  \label{proposition:simultaneous-chaining-true-dependency-graph}
  Let $X\in\bnd(\mathcal{E})$ and $i\in\npredset(\varphi_X)$ such that $\pred{\varphi_X}{i}=X$.
  Let $G=(V,\regularE)$ be a true-dependency graph for~$\mathcal{E}$, and let
  $I\subseteq\pviPositions{X}{\data{\varphi_X}{i}}{\varphi_X}$.
  Then $\chainedG$ is a true-dependency graph for~$\unfoldAll{\mathcal{E},X}{I}$.
\end{restatable}

\newcommand{\proofpropproofgraph}{%
  \begin{proof}
    \setlength{\parindent}{0pt}
    Since we are reasoning in the context of two different graphs with identical sets of vertices but different sets of edges, for $u\in V$ we write $u^\bullet$ for the set of successors of $u$ w.r.t.\ $\regularE$ and $u^\circ$ for the set of successors of $u$ w.r.t.\ $\regularE\cup\tilE{}$. Note that for all $u\in V$, $u^\bullet\subseteq u^\circ$.
    Fix some $(\depgraphX,\vec{v})\in V$.
    If $\depgraphX\neq X$, then the right-hand side is the same for $\mathcal{E}$ and~$\unfoldAll{\mathcal{E},X}{I}$. Then by monotonicity, for every data environment $\delta$,
    $\dbrack{\varphi_\depgraphX}((\depgraphX,\vec{v})^\bullet)_{\true}\delta[\vec{v}/\vec{d}_\depgraphX]
    \implies
    \dbrack{\varphi_\depgraphX}((\depgraphX,\vec{v})^\circ)_{\true}\delta[\vec{v}/\vec{d}_\depgraphX]$.
    Next, consider when $\depgraphX=X$.

    \noindent We first prove that for all data environments $\delta'$ and all $j\in I$, we have:
    \[
      \tag{$*$}\label{eqn:step-simultaneous}
      \dbrack{X(\data{\varphi_X}{j})}((X,\vec{v})^\bullet)_{\true}\delta'
      \implies
      \dbrack{\varphi_X\fvsub{\data{\varphi_X}{i}}}((X,\vec{v})^\circ)_{\true}\delta'.
    \]
    Let $\delta'$ be a data environment, $j\in I$, and assume $\dbrack{X(\data{\varphi_X}{j})}((X,\vec{v})^\bullet)_{\true}\delta'$ holds true. Since $j\in I$, we have $\pred{\varphi_X}{j}=X$ and $\data{\varphi_X}{j}=\data{\varphi_X}{i}$. Then also $(X,\dbrack{\data{\varphi_X}{i}}\delta')\in (X,\vec{v})^\bullet$ and so $\dbrack{\data{\varphi_X}{i}}\delta'\in\values((X,\vec{v}),j,\regularE)$. Define $\vec{u}=\dbrack{\data{\varphi_X}{i}}\delta'$.

    Since $(X,\vec{u})\in (X,\vec{v})^\bullet$, we may conclude that $(X,\vec{u})\in V$, and since $G$ is a true-dependency graph, we find that also $\dbrack{\varphi_X}((X,\vec{u})^\bullet)_{\true}\delta'[\vec{u}/\vec{d}_X]$ holds true. Moreover, because we have $\vec{u}=\dbrack{\data{\varphi_X}{i}}\delta'$, by \cref{lemma:from-delta-move-free-var-in}, we find that also $\dbrack{\varphi_X\fvsub{\data{\varphi_X}{i}}}((X,\vec{u})^\bullet)_{\true}\delta'$ holds true.

    From $\vec{u}\in\values((X,\vec{v}),j,\regularE)$ and $(X,\vec{v})\in V$ it follows, by definition of $\tilE[j]$ and since $\tilE[j]=\tilE{}$, that $(X,\vec{u})^\bullet\subseteq(X,\vec{v})^\circ$. So, by monotonicity, it must be the case that $\dbrack{\varphi_X\fvsub{\data{\varphi_X}{i}}}((X,\vec{v})^\circ)_{\true}\delta'$ holds true. Hence, we can conclude that~\eqref{eqn:step-simultaneous} holds true.

    Let $\delta$ be a data environment. We now show that $\dbrack{\varphi_X\isubsim{j}{\varphi_X\fvsub{\data{\varphi_X}{i}}}{I}}((X,\vec{v})^\circ)_{\true}\delta[\vec{v}/\vec{d}_X]$ holds true. Since $(X,\vec{v}) \in V$ and $G$ is a true-dependency graph, $\dbrack{\varphi_X}((X,\vec{v})^\bullet)_{\true}\delta[\vec{v}/\vec{d}_X]$ must hold true.
    But then we find that by~\cref{lemma:simultaneous-mapsto-identity} also $\dbrack{\varphi_X\isubsim{i}{X(\data{\varphi_X}{i})}{I}}( (X,\vec{v})^\bullet)_{\true}\delta[\vec{v}/\vec{d}_X]$ must be true.
    Combining \cref{lemma:simultaneous-substitution-monotonicity} and~\eqref{eqn:step-simultaneous}, we obtain:
    \[
      \begin{array}{l}
        \dbrack{\varphi_X\isubsim{j}{X(\data{\varphi_X}{i})}{I}}((X,\vec{v})^\bullet)_{\true}\delta[\vec{v}/\vec{d}_X] \\
        \implies
        \dbrack{\varphi_X\isubsim{j}{\varphi_X\fvsub{\data{\varphi_X}{i}}}{I}}((X,\vec{v})^\circ)_{\true}\delta[\vec{v}/\vec{d}_X]
      \end{array}
    \]
    So, we may conclude that $\dbrack{\varphi_X\isubsim{j}{\varphi_X\fvsub{\data{\varphi_X}{i}}}{I}}((X,\vec{v})^\circ)_{\true}\delta[\vec{v}/\vec{d}_X]$ holds true. Hence, $\simultaneousChainedG$ is a true-dependency graph for~$\unfoldAll{\mathcal{E},X}{I}$.
  \end{proof}
}
\ifapxpropproofgraph
\else
\proofpropproofgraph
\fi

\begin{restatable}{theorem}{simultaneousChainingProofGraph}
  \label{theorem:proof-graph-multiple-equation-chaining}
  Let $X\in\bnd(\mathcal{E})$, $i\in\npredset(\varphi_X)$ such that $\pred{\varphi_X}{i}=X$, and\iffull{} such that \fi
  $I\subseteq\pviPositions{X}{\data{\varphi_X}{i}}{\varphi_X}$.
  Let $G=(V,\regularE)$ be a proof graph for~$\mathcal{E}$.
  Then $\chainedG$ is a proof graph for~$\unfoldAll{\mathcal{E},X}{I}$.
\end{restatable}
\newcommand{\proofthmproofgraph}{%
  \begin{proof}
    Since $G$ is a proof graph for~$\mathcal{E}$, it is also a true-dependency graph. By \cref{proposition:simultaneous-chaining-true-dependency-graph}, $\simultaneousChainedG$ is a true-dependency graph for~$\unfoldAll{\mathcal{E},X}{I}$. Hence, it suffices to argue that all infinite paths in $\simultaneousChainedG$ are $\nu$-dominated. Let $\pi'$ be an arbitrary infinite path in $\simultaneousChainedG$. Towards a contradiction, assume $\pi'$ is not $\nu$-dominated. Then there are infinitely many subpaths $x\,y$ of length two that appear on $\pi'$ and for which $(x,y)\in\tilE{}\setminus\regularE$, for if there are only finitely many such paths, there is an infinite suffix of $\pi'$ that is also a path in $G$, which we know to be $\nu$-dominated, contradicting the assumption that $\pi'$ is not $\nu$-dominated.

    Consider such a subpath $x\,y$ in $\simultaneousChainedG$. By definition of $\tilE{}$, there is some $j\in I$ such that $(x,y)\in\tilE[j]$. By definition of $\tilE[j]$, $x$ must be of the shape $(X,\vec{v})$ while $y$ must be of the shape $(Y,\vec{w})$. Moreover, there must be some $\vec{u}$ such that $\vec{u}\in\values((X,\vec{v}),j,\regularE)$ and $((X,\vec{u}),(Y,\vec{w}))\in\regularE$. Let $\vec{u}$ be such. By definition of $\values$, we have $\lb{(X,\vec{v}),(X,\vec{u})}\in\regularE$. Hence, there is a path $(X,\vec{v})\,(X,\vec{u})\,(Y,\vec{w})$ of length three in $G$. Furthermore, the fixpoint equations visited by this path are the same as those visited by the subpath $(X,\vec{v})\,(Y,\vec{w})$ in $\simultaneousChainedG$. So, for every subpath $x\,y$ of length two in $\simultaneousChainedG$ that is not in $G$, there is a path of length three from $x$ to $y$ in $G$ that furthermore visits the same fixpoint equations.

    But this means that we can replace each subpath $x\,y$ on $\pi'$, for which $(x,y)\in\tilE{}\setminus\regularE$, by another subpath $x\,z\,y$ such that $(x,z)\in\regularE$ and $(z,y)\in\regularE$, obtaining a path $\pi$ in $G$. Since $\pi$ and $\pi'$ visit the same fixpoint equations and $\pi'$ is not $\nu$-dominated, also $\pi$ cannot be $\nu$-dominated. But this contradicts that $G$ is a proof graph. So $\pi'$ is $\nu$-dominated and therefore $\simultaneousChainedG$ is a proof graph for $\unfoldAll{\mathcal{E},X}{I}$.
  \end{proof}
}
\ifapxthmproofgraph
\else
\proofthmproofgraph
\fi
\iffull
\begin{figure}[tbp]
  \centering
  {
    \begin{tikzpicture}[node distance=2cm, ->, >=stealth]
      \node (G) at (4,3) {$G$:};
      \node (\chainedG) at (4,1) {$\chainedG$:};

      \node (k_n) at (5,3) {$(X,\vec{v}$)};
      \node (k_n1) at (10,3) {$(Y,\vec{w})$};
      \node (tilV) at (7.5,3) {$(X,\vec{u})$};
      \draw[->] (k_n) -- node[midway, above] {$\in \regularE$} (tilV);
      \draw[->] (tilV) -- node[midway, above] {$\in \regularE$} (k_n1);

      \node (k_n') at (5,1) {$(X,\vec{v})$};
      \node (k_n1') at (10,1) {$(Y,\vec{w})$};
      \draw[->] (k_n') -- node[midway, above] {$\in \tilE \setminus \regularE$} (k_n1');

      \draw[-,dashed] (k_n) -- (k_n');
      \draw[-,dashed] (k_n1) -- (k_n1');

    \end{tikzpicture}
  }
  \caption{Mirroring segments of an infinite path from $\chainedG$ in $G$}
  \label{fig:viz-proof-graph}
\end{figure}
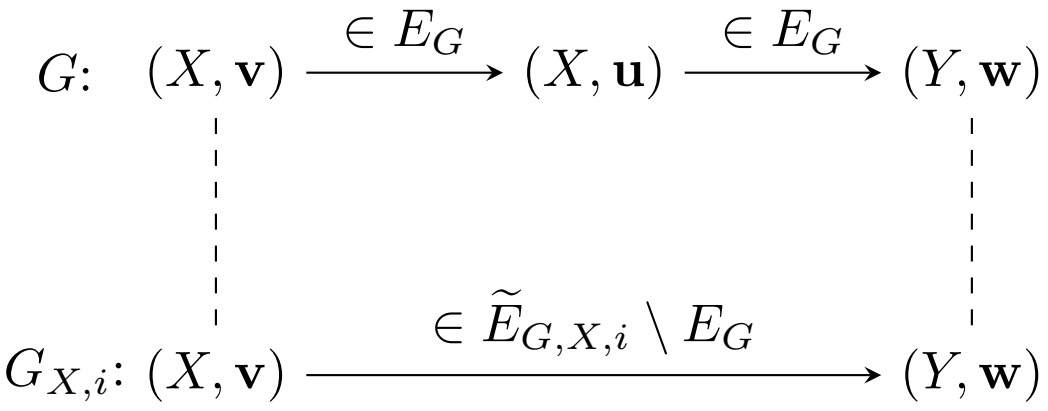
\fi
\Cref{theorem:proof-graph-multiple-equation-chaining} also reveals that the size increase of the certificate for a solution to a PBES after chaining, \emph{i.e.}~its proof or refutation graph, is limited. This does not mean that the space in which such a certificate can be found is limited similarly.
We next explore the effects chaining may have on the relevant set of predicate variable instances and their inter-dependencies that are needed to solve a PBES.
For this, we study the impact chaining has on a relevancy graph.
In~\cref{theorem:relevancy-graph-srf}, we give an upper bound on the size of the set of relevant edges for a chained PBES in standard recursive form.
Unfortunately, such bounds cannot be established for PBESs in general as the example below shows.

\begin{example}\label{example:uncontrolled_blow_up}
  Consider the equation system~$\mathcal{E}$ below.
  Notice that $(X,2)$ is the only possibly relevant vertex for $(X,1)$.
  \[
    \begin{split}
      \mu X(n\sort N) = (&n = 1) \implies (X(2) \land (X(2*n) \lor \exists m\sort N. X(m+3))) \\
      &\land (n = 2) \implies X(1)
    \end{split}
  \]
  \iffsen
\else
  After chaining the first PVI, \emph{i.e.},~$X(2)$, we get the following system.
  \[
    \begin{split}
      \mu X(n\sort N) = (&n = 1) \implies (X(1) \land (X(2*n) \lor \exists m\sort N. X(m+3))) \\
      &\land (n = 2) \implies X(1)
    \end{split}
  \]
  \fi
  Vertices that were previously unreachable in the relevancy graph for $\mathcal{E}$ suddenly become relevant for $(X,1)$, resulting in an infinite-sized relevancy graph.
  \begin{center}
    \begin{tikzpicture}[node distance=2cm, ->, >=Stealth]
      \node (x1) at (0,0) {$(X,1)$};
      \node (x2) at (2,0) {$(X,2)$};
      \node (x3) at (4,0) {$(X,3)$};
      \node (x4) at (6,0) {$(X,4)$};
      \node (x5) at (8,0) {$(X,5)$};
      \node (x6) at (10,0) {$\dots$};

      \draw[->] (x1) to (x2);
      \draw[->,loop left] (x1) to (x1);
      \draw[->, bend right=15] (x2) to (x1);
      \draw[->, bend right=15] (x1) to (x3);
      \draw[->, bend right=15] (x1) to (x4);
      \draw[->, bend right=15] (x1) to (x5);
      \draw[->, bend right=15] (x1) to (x6);
    \end{tikzpicture}

  \end{center}
\end{example}
When considering a PBES in standard recursive form, we can guarantee no new nodes are introduced in the relevancy graph of a chained PBES.
Before we consider the main theorem, we introduce two lemmata. We use the first to prove the second, which is then used for the proof of the main theorem. The first lemma shows the direct correspondence between relevant edges and the semantics of the equation system.

\begin{restatable}{lemma}{srfrelevantiff}
  \label{lemma:srf-relevant-iff}
  Suppose that~$\mathcal{E}$ is in SRF.
  Let $G=(V,\regularE)$ be a relevancy graph for~$\mathcal{E}$,
  and let $(X,\vec{v}),(Y,\vec{w}) \in V$. Then $(Y,\vec{w}) \in (X,\vec{v})^\bullet$ iff
  for some $j\in J_X, v_j \in \mathbb{E}_j$ we have that $\dbrack{f_j(\vec{d},e_j)}\delta[\vec{v}/\vec{d}][v_j/e_j]$ holds, $Y = X_j$, and $\vec{w} = \dbrack{\vec{g}_j(\vec{d},e_j)}\delta[\vec{v}/\vec{d}][v_j/e_j]$.
\end{restatable}
\noindent\ignorespacesafterend
The following lemma is the basis for establishing that chaining does not add new nodes to the relevancy graph of the PBES when it is in SRF.
Additionally, for each node associated with a chained equation, a node is not relevant if that node was not already relevant and that node is not relevant for some relevant evaluation of the PVI at position $i$.

\begin{restatable}{lemma}{simnonewrelevanctedges}
  \label{lemma:no-new-relevant-edges}
  Suppose that~$\mathcal{E}$ is in SRF. Let $X \in \bnd(\mathcal{E})$, $i \in \npredset(\varphi_X)$ be such that $\pred{\varphi_X}{i} = X$, and $I \subseteq \pviPositions{X}{\data{\varphi_X}{i}}{\varphi_X}$.
  Let $G=(V,\regularE)$ be a relevancy graph for~$\mathcal{E}$, and let $(X,\vec{v}) \in V$.
  Let $(Y,\vec{w}) \in \sgn(\mathcal{E}) \setminus (X,\vec{v})^\bullet$ and assume that $(X,\dbrack{\data{\varphi_X}{i}}\delta[\vec{v}/\vec{d}_X])\in (X,\vec{v})^\bullet$ implies $(Y,\vec{w}) \notin (X,\dbrack{\data{\varphi_X}{i}}\delta[\vec{v}/\vec{d}_X])^\bullet$ for all $\delta$.
  \iffsen
\else
  Then for all $\eta,\delta$, it holds that $\dbrack{\unfoldAll{\varphi_X}{I}}\eta\pvisub{\top}{Y(\vec{w})}\delta[\vec{v}/\vec{d}_X]\allowbreak = \dbrack{\unfoldAll{\varphi_X}{I}}\eta\pvisub{\bot}{Y(\vec{w})}\delta[\vec{v}/\vec{d}_X]$.
  \fi

\end{restatable}
\noindent\ignorespacesafterend
\iffsen
\else
The theorem below provides an upper bound on the size of the relevancy graph of a chained PBES in SRF, compared to the relevancy graph of the original PBES.
Notice that in this theorem, the set $\tilE$, see~\cref{def:values-egxi}, is the same as the one used in the proof graph theorem.
\fi

\newcommand{\chainedGSRF}{G'}
\newcommand{\chainedV}{V'}
\newcommand{\chainedE}{E'}
\newif\ifapxrelevancygraphsrf
\apxrelevancygraphsrffalse
\newif\ifapxsimrelevancygraphsrf
\iffsen
\else
\apxsimrelevancygraphsrffalse
\fi
\begin{restatable}{theorem}{simrelevancygraphsrf}
  \label{theorem:relevancy-graph-srf}
  Let~$\mathcal{E}$ be an equation system in SRF.
  Let $X \in \bnd(\mathcal{E})$, $i \in \npredset(\varphi_X)$ such that $\pred{\varphi_X}{i} = X$, and $I \subseteq \pviPositions{X}{\data{\varphi_X}{i}}{\varphi_X}$.
  Let $G=(V,\regularE)$ be a relevancy graph for~$v \in \sgn(\mathcal{E})$.
  Then there is a relevancy graph $\chainedGSRF=(V,\chainedE)$ for $\unfoldAll{\mathcal{E},X}{I}$.
  Specifically, $\chainedE \subseteq \regularE \cup \tilE$.
\end{restatable}
\newcommand{\xprime}{Z}
\newcommand{\proofsimrelevancygraphsrf}{%
  \begin{proof}
    \setlength{\parindent}{0pt}
    Abbreviate $\unfoldAll{\mathcal{E},X}{I}$ by $\mathcal{E}'$.
    We use $\varphi_\xprime$ to refer to the right-hand side of equation $\xprime$ in~$\mathcal{E}$, and $\varphi_\xprime'$ to refer to the right-hand side of equation $\xprime$ in $\mathcal{E}'$.

    Pick some arbitrary pair $(\xprime,\vec{v}) \in V$ and $(Y,\vec{w}) \in \sgn(\mathcal{E})$ such that there is some $\eta,\delta$ such that $\dbrack{\varphi_{\xprime}'}\eta\pvisub{\top}{Y(\vec{w})}\delta[\vec{v}/\vec{d}_{\xprime}] \neq \dbrack{\varphi_{\xprime}'}\eta\pvisub{\bot}{Y(\vec{w})}\delta[\vec{v}/\vec{d}_{\xprime}]$.
    We show that $(V, \chainedE)$ is a valid relevancy graph by showing that $(Y,\vec{w}) \in V$. 
    Specifically, we show that $\lb{(\xprime,\vec{v}),(Y,\vec{w})} \in \regularE \cup \tilE$, which is sufficient since $\regularE \cup \tilE \subseteq V \times V$.
    Indeed we show that $\lb{(\xprime,\vec{v}),(Y,\vec{w})}\notin \regularE$ implies $\lb{(\xprime,\vec{v}),(Y,\vec{w})}\in \tilE$.
    Assume $\lb{(\xprime,\vec{v}),(Y,\vec{w})}\notin \regularE$. Then it must be the case that for all $\eta,\delta$, $\dbrack{\varphi_{\xprime}}\eta\pvisub{\top}{Y(\vec{w})}\delta[\vec{v}/\vec{d}_{\xprime}] = \dbrack{\varphi_{\xprime}}\eta\pvisub{\bot}{Y(\vec{w})}\delta[\vec{v}/\vec{d}_{\xprime}]$.
    It must be that $\xprime = X$, since $\varphi_{\xprime}' \neq \varphi_{\xprime}$ and for all $Q\neq X$, we know $\varphi_{Q}' = \varphi_{Q}$.
    Therefore $\varphi_{\xprime}' = \unfoldAll{\varphi_X}{I}$.
    Also by assumption $(Y,\vec{w}) \in \sgn(\mathcal{E})\setminus (X,\vec{v})^\bullet$. Therefore by~\cref{lemma:no-new-relevant-edges}, it must be that there is some $\delta'$ with $\vec{u} \seteq \dbrack{\data{\varphi}{i}}\delta'[\vec{v}/\vec{d}_X]$ such that $(X,\vec{u})\in (X,\vec{v})^\bullet$ and $(Y,\vec{w}) \in (X,\vec{u})^\bullet$.
    Therefore, $\lb{(X,\vec{v}),(Y,\vec{w})}\in \tilE$.
  \end{proof}
}
\ifapxsimrelevancygraphsrf
\else
\proofsimrelevancygraphsrf
\fi
\begin{myomit}
\newcommand{\proofrelevancygraphsrf}{%
  \begin{proof}
    \newcommand{\xprime}{Z}
    \setlength{\parindent}{0pt}
    We use $\varphi_\xprime$ to refer to the right-hand side of equation $\xprime$ in~$\mathcal{E}$, and $\varphi_\xprime'$ to refer to the right-hand side of equation $\xprime$ in $\mathcal{E}'$.

    Pick some $\lb{(\xprime,\vec{v}),(Y,\vec{w})}\in \chainedE$.
    We need to show that $\lb{(\xprime,\vec{v}),(Y,\vec{w})}\in \regularE \cup \tilE$.
    Indeed we show that $\lb{(\xprime,\vec{v}),(Y,\vec{w})}\notin \regularE$ implies $\lb{(\xprime,\vec{v}),(Y,\vec{w})}\in \tilE$.
    Assume $\lb{(\xprime,\vec{v}),(Y,\vec{w})}\notin \regularE$. Then for all $\eta,\delta$, $\dbrack{\varphi_{\xprime}}\eta\pvisub{\top}{Y(\vec{w})}\delta[\vec{v}/\vec{d}_{\xprime}] = \dbrack{\varphi_{\xprime}}\eta\pvisub{\bot}{Y(\vec{w})}\delta[\vec{v}/\vec{d}_{\xprime}]$.
    By assumption, there is some $\eta,\delta$ such that $\dbrack{\varphi_{\xprime}'}\eta\pvisub{\top}{Y(\vec{w})}\delta[\vec{v}/\vec{d}_{\xprime}] \neq \dbrack{\varphi_{\xprime}'}\eta\pvisub{\bot}{Y(\vec{w})}\delta[\vec{v}/\vec{d}_{\xprime}]$.
    It must be that $\xprime = X$, since $\varphi_{\xprime}' \neq \varphi_{\xprime}$ and for all $Q\neq X$, we know $\varphi_{Q}' = \varphi_{Q}$.
    Since  $\lb{(X,\vec{v}),(Y,\vec{w})}\in \chainedE$, we know that $\dbrack{\unfold{\varphi_X}{i}}\eta\pvisub{\top}{Y(\vec{w})}\delta[\vec{v}/\vec{d}_X] \neq \dbrack{\unfold{\varphi_X}{i}}\eta\pvisub{\bot}{Y(\vec{w})}\delta[\vec{v}/\vec{d}_X]$.
    Also by assumption $(Y,\vec{w}) \notin (X,\vec{v})^\bullet$. Therefore by~\cref{lemma:no-new-relevant-edges}, it must be that there is some $\delta'$ with $\vec{u} \seteq \dbrack{\data{\varphi}{i}}\delta'[\vec{v}/\vec{d}_X]$ such that $(X,\vec{u})\in (X,\vec{v})^\bullet$ and $(Y,\vec{w}) \in (X,\vec{u})^\bullet$.
    Therefore, $\lb{(X,\vec{v}),(Y,\vec{w})}\in \tilE$.
    %
  \end{proof}
}
\ifapxrelevancygraphsrf
\else
\proofrelevancygraphsrf
\fi
\end{myomit}
It follows from the theorem that the relevancy graph of a chained PBES $\chainedG$ can only add new dependencies between vertices that were already present in the relevancy graph of the original PBES; no new vertices are introduced.
The worst-case blow-up is therefore trivially limited to be at most quadratic in the number of vertices of the relevancy graph underlying the original PBES.
While~\cref{theorem:relevancy-graph-srf} offers insights into the maximal increase that can be expected when chaining a PBES in SRF, we expect that, as, \emph{e.g.}, illustrated in~\cref{example:less-dependency-after-chaining}, chaining may actually result in a size reduction of the relevancy graph, both in the number of vertices and the number of edges.







\section{Heuristics for Effective Chaining}
\label{sec:heuristics}

\cref{example:uncontrolled_blow_up} shows that chaining may have a dramatic effect on the blow-up of the relevancy graph underlying a PBES, whereas~\cref{theorem:relevancy-graph-srf} shows this effect is limited in case the PBES is in SRF.
In order to effectively exploit the power of chaining, we next study cases under which a blow-up typically occurs and propose a set of heuristics to help avoid adverse effects of chaining.
Based on these observations, we devise a heurstics-informed technique called \emph{selective chaining}, that underlies an algorithm, presented at the end of this section, that can be used to effectively use chaining to simplify a PBES.



\paragraph{Avoiding Quantifier-bound PVI (H1).}
Chaining PVIs that occur in the scope of a quantifier can result in a significant increase in the number of edges in the relevancy graph underlying a chained PBES.
We illustrate this using the following equation system.
\[
  \begin{split}
    \mu X(n: N) &= \forall m\sort N. (n \approx 0 \land m > 0 \land m < 50) \implies  X(m) \\
    &\land (n > 0 \land n < 50) \implies X(50)\\
    &\land \forall m\sort N. (n \approx 50 \land m \geq 51 \land m \leq 53) \implies  X(m) \\
  \end{split}
\]
A relevancy graph $G$ of the above equation system, containing $X(0)$, is shown below (left).
\iffsen
\else
\begin{center}
  {\centering
    \begin{minipage}[htbp]{.49\linewidth}
      \centering
      \begin{tikzpicture}[node distance=2cm, ->, >=stealth]
        \node (xi) at (-2,2) {$(X,0)$};
        \node (x0) at (2,2) {$(X,50)$};
        \node (x1) at (0,3) {$(X,1)$};
        \node (x2) at (0,2) {$(X,2)$};
        \node (x3) at (0,1) {$(X,3)$};
        \node (x4) at (0,0) {$\vdots$};
        \node (y1) at (4,3) {$(X,51)$};
        \node (y2) at (4,2) {$(X,52)$};
        \node (y3) at (4,1) {$(X,53)$};

        \draw[->] (xi) -- (x1);
        \draw[->] (xi) -- (x2);
        \draw[->] (xi) -- (x3);
        \draw[->, dashed] (xi) -- (x4);
        \draw[->] (x1) -- (x0);
        \draw[->] (x2) -- (x0);
        \draw[->] (x3) -- (x0);
        \draw[->, dashed] (x4) -- (x0);
        \draw[->] (x0) -- (y1);
        \draw[->] (x0) -- (y2);
        \draw[->] (x0) -- (y3);
      \end{tikzpicture}
    \end{minipage}
    \begin{minipage}[htbp]{.49\linewidth}
      \centering
      \begin{tikzpicture}[node distance=2cm, ->, >=stealth]
        \node (xi) at (-2,2) {$(X,0)$};
        \node (x1) at (0,3) {$(X,1)$};
        \node (x2) at (0,2) {$(X,2)$};
        \node (x3) at (0,1) {$(X,3)$};
        \node (x4) at (0,0) {$\vdots$};
        \node (y1) at (3,3) {$(X,51)$};
        \node (y2) at (3,2) {$(X,52)$};
        \node (y3) at (3,1) {$(X,53)$};


        \draw[->] (xi) -- (x1);
        \draw[->] (xi) -- (x2);
        \draw[->] (xi) -- (x3);
        \draw[->, dashed] (xi) -- (x4);

        \draw[->, dashed] (x4) -- (y1);
        \draw[->, dashed] (x4) -- (y2);
        \draw[->, dashed] (x4) -- (y3);

        \draw[->, bend left=20] (x3) to (y1);
        \draw[->] (x3) -- (y2);
        \draw[->] (x3) -- (y3);

        \draw[->, bend left=20] (x2) to (y1);
        \draw[->] (x2) to (y2);
        \draw[->, bend right=20] (x2) to (y3);

        \draw[->] (x1) -- (y1);
        \draw[->] (x1) -- (y2);
        \draw[->, bend right=20] (x1) to (y3);

      \end{tikzpicture}
    \end{minipage}
  }
\end{center}
\fi
When we chain with $I=\lc{2}$, PVI $X(50)$ is replaced by $\forall m\sort N. (m \geq 51 \land m \leq 53) \implies X(m)$. Now $|\regularE \cup E_{G,X,2}|\ = 196$ while $|\regularE|\ = 101$.
Whether or not a quantifier gives rise to a blow-up cannot cheaply be computed.
Our \emph{avoiding quantifier-scoped PVI} heuristic thus prevents chaining when it introduces a PVI in the unfolded expression that is in the scope of a quantifier.

\paragraph{Bound PVI Growth per Chaining Action (H2).}

The number of additional PVIs that are introduced in the chained PBES is always limited but as is clear from~\cref{example:uncontrolled_blow_up}, the relevancy graph of the chained PBES may become problematically large.
Observe, however, that for PBESs in SRF, the size increase is limited in case the PVIs that are introduced due to chaining are not within the scope of a quantifier.
In that case, when chaining introduces $k$ such PVIs, the set of edges of the relevancy graph underlying the chained PBES can grow no more than linearly in $k$.
To see this, consider the following PBES.
\[\nu X(n\sort N) = X(2) \land (n \approx 1 \implies X(n+1)) \land (n \approx 2 \implies X(n+1))\]
The relevancy graph for $(X,1)$, underlying the PBES is depicted below (left).
Observe that the PBES has~3 PVIs in total, two of which are syntactically identical.
Applying chaining to the $X(2)$ PVI will result in~2 new PVIs, syntactically.
Chaining applied to the $X(n+1)$ PVIs results in the following PBES (after arithmetic simplification):
\[\nu X(n\sort N) = X(2) \land
  \begin{array}[t]{l}
    (n \approx 1 \implies (X(2) \land (n \approx 0 \implies X(n+2)) \land (n \approx 1 \implies X(n+2)))) \land {} \\
    (n \approx 2 \implies (X(2) \land (n \approx 0 \implies X(n+2)) \land (n \approx 1 \implies X(n+2))))
  \end{array}
\]
Both occurrences of PVI $X(n+1)$ are (syntactically) replaced by 3~new PVIs.
Since these new PVIs are not within the scope of a quantifier, their impact on the size increase of the relevancy graph of the chained PBES is limited.
In fact, for the example PBES, only one additional edge is added to the relevancy graph underlying it, see the graph depicted on the right.
\begin{center}
  \centering
  \begin{tikzpicture}[node distance=2cm, ->, >=Stealth]
    \node (x1) at (0,0) {$(X,1)$};
    \node (x2) at (1.5,0) {$(X,2)$};
    \node (x3) at (3,0) {$(X,3)$};

    \draw[->] (x1) edge (x2);
    \draw[->,loop above] (x2) edge (x2);
    \draw[->,bend left] (x3) edge (x2);
    \draw[->,bend left] (x2) edge (x3);

    \node (x21) at (5.5,0) {$(X,1)$};
    \node (x22) at (7,0) {$(X,2)$};
    \node (x23) at (8.5,0) {$(X,3)$};

    \draw[->] (x21) edge (x22);
    \draw[->,loop above] (x22) edge (x22);
    \draw[->,bend left] (x22) edge (x23);
    \draw[->,bend left] (x23) edge (x22);
    \draw[->, bend right=25] (x21.south) to (x23.south);
    %
  \end{tikzpicture}
\end{center}
The effect on the relevancy graph can be approximated syntactically by applying stronger forms of logical simplification to the PBES obtained after chaining.
For instance, a powerful rewriter might simplify the resulting PBES to:
\[\nu X(n\sort N) = X(2) \land (n \approx 1 \implies X(n+2))
\]
Our \emph{bound PVI growth per chaining action} heuristic takes the increase in the number of PVIs obtained after chaining \emph{and} rewriting into account and puts a limit on this increase, thus guaranteeing that the blow-up is always limited for PBESs in SRF if the newly introduced PVI are not in the scope of quantifiers.

\paragraph{Transforming to SRF before Chaining (H3).}
The SRF normal form is essential when relying on algorithms that use decision diagrams to compactly represent, among others, the relevancy graph and solve a PBES symbolically.
Due to the nature of the transformation of chaining, a PVI in a non-SRF PBES may be turned in a formula that no longer contains any PVI.
For instance, a PVI $X(\vec{e})$ in a clause $(f \land X(\vec{e})) \lor \bigvee \ldots$ in such a PBES could be replaced by $\top$ as a result of chaining.
When subsequently converting the PBES to SRF, when formula $f$ contains no PVIs, it gets distributed across the disjuncts when transforming the formula to SRF.
This can cause additional overhead when solving the PBES.
Our \emph{transforming to SRF before chaining} heuristic counters this effect by converting a PBES to SRF prior to chaining whenever the solver that is used requires the PBES to be in SRF.
By~\cref{theorem:relevancy-graph-srf}, we are, moreover, guaranteed more stricter bounds on the size increase of the relevancy graph caused by chaining.

\paragraph{Limit Size of Unfolded Formula (H4).}
Even with the heuristics discussed above, chaining can still yield a problematic increase in the size of a PBES, reflecting a potential increase in the computational effort required to solve the PBES.
We illustrate this using a very simple PBES, in which a formula $\psi$, possibly containing parameter $n$, occurs once:
\[ \nu X(n\sort N) = 
  (n \leq 4 \implies X(3)) \land (n \geq 2 \implies X(3)) \land (n \approx 3 \implies (X(3) \land \fp{\psi}{n}))
\]
When applying chaining to $X(3)$, all three occurrences of $X(3)$ are replaced by expressions in which formula $\psi$ appears syntactically (after minor logical and arithmetic simplification):
\[ \nu X(n\sort N) = 
  n \leq 4 \implies (X(3) \land \psi\fvsub{3}) \land (n \geq 2 \implies (X(3) \land \psi\fvsub{3})) \land (n \approx 3 \implies (X(3) \land \psi\fvsub{3}))
\]
When $\psi$ is (syntactically) is large, this duplication may incur a significant penalty on the computational effort involved in solving a PBES.
Our \emph{limit the size of unfolded formula} heuristic puts a limit on the size increase of the formula: we will consider the maximum allowed size of the unfolded formula size to be the size of the original PVI in textual form, multiplied with a predetermined factor~(1.5, 2, \ldots).

\newcommand{\varstyle}[1]{\textsf{#1}}
\newcommand{\todorhs}{\varstyle{PVIs}}
\newcommand{\varX}{{X}}
\newcommand{\varD}{{e}}
\newcommand{\procChain}{Chain}
\newcommand{\gausselimination}{GaussElm}
\newcommand{\dataX}{d}
\newcommand{\DataX}{D}
\newcommand{\sigmaX}{\sigma}
\newcommand{\todochain}{\varstyle{todo}}
\newcommand{\visitedchain}{\varstyle{done}}
\newcommand{\curargs}{\varstyle{pvi\_arg}}
\newcommand{\problemsize}{\varstyle{max\_work\_stack}}
\begin{algorithm}[tbp]
  \centering
  \begin{minipage}[t]{.8\linewidth}
    \caption{Repeatedly apply selective chaining for an equation}\label{alg:reducerhs}
    \begin{algorithmic}[1]
      \Procedure{ReduceRHS}{$\sigma X(d\sort D) = \varphi_X$}
      \State $\todorhs \gets \lc{\varD \mid X(\varD) \in \pvi(\varphi_X)}$
      \While{$\todorhs \neq \varnothing$}
      \State $\varD \gets \Call{pick}{\todorhs}$
      \State $\todorhs \gets \todorhs \setminus \lc{\varD}$
      \State $\todochain \gets \lc{\varD}$
      \State $\visitedchain \gets \varnothing$
      \While {$\left|\todochain\right| \neq \varnothing \land \left|\visitedchain\right| + \left|\todochain\right| < \problemsize$}
      \State $\curargs \gets \Call{pick}{\todochain}$
      \State $\todochain \gets \todochain \setminus \lc{\curargs}$
      \State $\visitedchain \gets \visitedchain \cup \lc{\curargs}$
      \State $\psi \gets {\varphi_X\fvsub{\curargs}}$
      \If {$\npred(\psi) \leq k \land \neg\Call{QuantPVI}{\psi} \land \neg\Call{TooBig}{\psi}$}
      \State $I \gets \pviPositions{X}{\small\curargs}{\varphi_X}$
      \State $\varphi_X \gets \varphi_X\isubsim{i}{\varphi_X\fvsub{\curargs}}{I}$  \Comment{$\unfoldAll{\varphi_X}{I}$}
      \State $\todochain \gets \todochain \cup \lc{\data{p}{} \mid p \in \pvi(\psi) \land \data{p}{} \notin \visitedchain}$
      \EndIf
      \EndWhile
      \EndWhile
      \EndProcedure
    \end{algorithmic}
  \end{minipage}
\end{algorithm}

\paragraph{The Selective Chaining Algorithm.}

The high-level procedure by which we apply selective chaining to a PBES works as follows.
Each equation is processed individually, from the highest rank to the lowest~(w.r.t.~$\triangleleft_{\mathcal{E}}$).
If the resulting right-hand side of an equation for $X$ is simple, \ie{} contains no PVI, we substitute the current equation into the other equations, replacing all PVIs with name identical to $X$ with the chained right-hand side of $X$~(\cf{} Lemma 19 and 18 of \cite{grooteParameterisedBooleanEquation2005}).
When working with PBESs that are in SRF, after this substitution we trigger a new transformation to SRF since the substitution can cause the right-hand sides to no longer be standard recursive clauses.

When processing an equation $(\sigma X(d\sort D) = \varphi_X)$, we loop through all PVIs for which its name is the same as the name of the equation we are considering, see Algorithm~\ref{alg:reducerhs}.
This algorithm works as follows.
It identifies the set of PVI arguments occurring in $\varphi_X$ and whose name is $X$.
These are subsequently processed one at a time:  \Call{pick}{} chooses a PVI argument $\varD$ from the set of $\todorhs$ to chain until there is no more PVI argument left.
We perform the chaining according to the second while-loop.
We first unfold $\varX(\varD)$, since $\varD$ is the only PVI argument in $\todochain$; the result is stored in $\psi$.
The procedure stops when one of enabled heuristics fails: the unfolded formula contains too many PVI~($\npred(\psi) > k$), the formula contains quantifier-scoped PVI~($\Call{QuantPVI}{\psi}$), or it exceeds the configured size bound~($\Call{TooBig}{\psi}$). If the formula is simple, chaining stops, otherwise, \Call{pick}{} selects one of its PVI arguments for the next step.

Termination of~\cref{alg:reducerhs} follows from the fact that the set $\todorhs$ is finite, and that we have a bound \problemsize{} while $\left|\visitedchain\right|$ increases every iteration~(line 11): it holds invariantly that all elements in \todochain{} and \visitedchain{} are disjoint sets, so \visitedchain{} gets a new element every iteration.
In our case studies in the next section, we repeatedly apply~\cref{alg:reducerhs} for an equation until we reach a convergence on the right-hand side of an equation.

\section{Practical Effective Application of Chaining}
\label{sec:practical_effectiveness}

We implemented selective chaining as a PBES transformation in a tool called~\pbeschain{} in the mCRL2 toolset~\cite{bunteMCRL2ToolsetAnalysing2019b}.
In this section we discuss the experiments we performed to assess the effectiveness of the transformation to help solve PBESs originating from real-life model checking problems.

\subsection{Case Studies}

The applicability of selective chaining highly depends on the model and property from which the PBES is generated. In models with substantial parallelism, interleaving typically results in equations that have multiple distinct successor PVIs, making chaining with a bounded PVI expansion harder. We see this when applying \pbeschain{} to, \emph{e.g.}, the DIRAC WMS~\cite{DBLP:journals/fgcs/RemenskaWVTB13} and ERTMS~\cite{bartholomeusModellingAnalysingERTMS2018} models.
In those cases, the time to perform chaining is neglible: it takes only a few seconds to perform the chaining, and since chaining is effectively not changing the PBES, the high solving times dominate.

When models contain choices that lead to paths where there is little choice in the next steps, selective chaining can often help. Such is for instance the case with control heavy applications. We show this in the following control heavy models. Different PBES solvers, using symbolic and explicit techniques, are applied to test how the effect of chaining changes for each type of solver. The explicit solver instantiates all reachable predicate variable instances, whereas a symbolic solver compactly represents sets of instantiations using decision diagrams.
Depending on the case study, using one of these solvers may be more optimal.
The first model is public, while the other two are confidential.

\newcommand{\casestudy}[1]{{[\texttt{C-#1}]}}
\begin{itemize}
  \item[\casestudy{1}] Prinses Marijke lock complex~\cite{grooteFormalSpecificationDesired2025}. On the model, 53 properties were verified. For our experiments, we also add 8 properties which were initially formed but disregarded by the authors because they did not hold or they were deemed not relevant. For our purposes, they provide additional realistic properties to benchmark our chaining heuristics. The range of time to check a requirement is very large, so we make use of log-log graphs for this case study. \hfill \textbf{(Symbolic)}
  \item[\casestudy{2}] Maeslant Storm Surge Barrier~\cite{beersCompleteFormalSpecification2026}. One of the three main components of the control software is modelled and verified using 40 properties. The PBESs for this case study contain more complex data expressions compared to the other case studies.\hfill \textbf{(Symbolic, Explicit)}
  \item[\casestudy{3}] Pneumatic Cylinder~\cite{stramagliaFormalVerificationIndustrial2022}. The model is derived from the Cordis SUITE, and 12 properties were verified on the model.\hfill \textbf{(Explicit)}
\end{itemize}
We compared how much time verifying each requirement took with and without adding \pbeschain{} as a preprocessing step.
The benchmarks were run on a machine with 4 x Intel Xeon 6136 CPUs and 3 TB of RAM on Ubuntu Server 24.04 LTS.
\iffull
The commands used can be found in Appendix~\ref{appendix:benchmarks}.
\else
\fi
We use scatterplots to visualize our experimental data. Each dot represents a property of the model.
We will use \casestudy{1} to refer to the first case study, and \casestudy{1-N} when $N$ is the limit of PVI expansion per chaining action as given by the heuristic (H2).
\iffull

\fi
When discussing individual heuristics, by default all other heuristics are enabled as well. Then, we can see if individual heuristic provide any benefit that is not provided by another heuristic. The settings are as follows: Bounded PVI Expansion = 1, SRF for symbolic cases, and Unfolded Formula Size Factor = 2 for \casestudy{2}.

\subsection{Applying Heuristics}
\paragraph{Avoiding Quantifier-scoped PVI (H1).}


We assess the effect of the heuristic that blocks quantifier-scoped PVIs to occur in the unfolding of a PVI. The PBESs for \casestudy{1} does not contain quantifiers, so we focus on~\casestudy{2}. For all but three requirements, the difference in the time to solve and the time to chain is at most a few percentage points. In three of the explicit solving cases, there is a more meaningful difference. The requirement \lstinline|2_devices-1_dock-2_door-dock_winch_one| requires\ slightly less time chaining (102s vs 121s) and is solved twice as fast (2740s vs 5664s) with the heuristic enabled. In this case, the heuristic does indeed block chaining for many PVIs. Thus the heuristic does not block any helpful chaining since it is never slower, and it does block some counterproductive chaining.

\paragraph{Bounded PVI expansion (H2).}

\begin{figure}[btp]
  \centering
  \begin{subfigure}[t]{0.49\linewidth}
    \includegraphics[width=\textwidth]{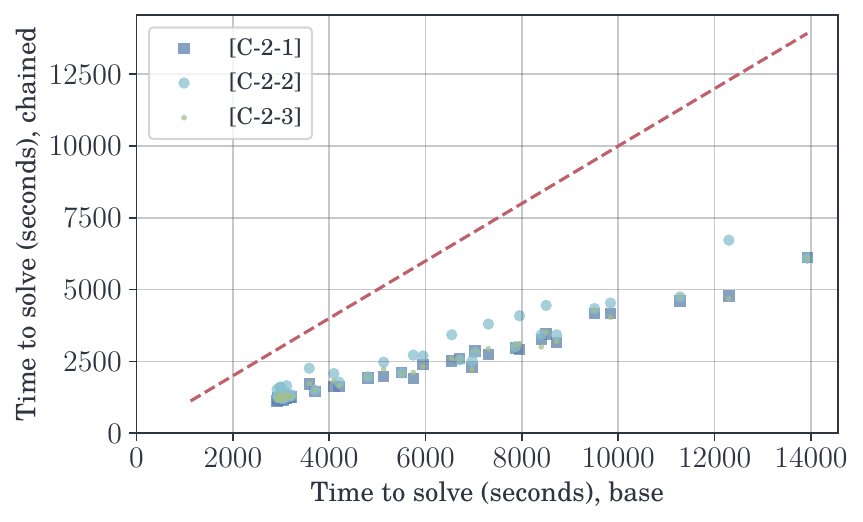}
    \caption{Comparison of time to solve PBES explicitly before and after chaining.}
    \label{fig:maeslant-maxpvi}
  \end{subfigure}
  \hfill
  \begin{subfigure}[t]{0.49\linewidth}
    \includegraphics[width=\textwidth]{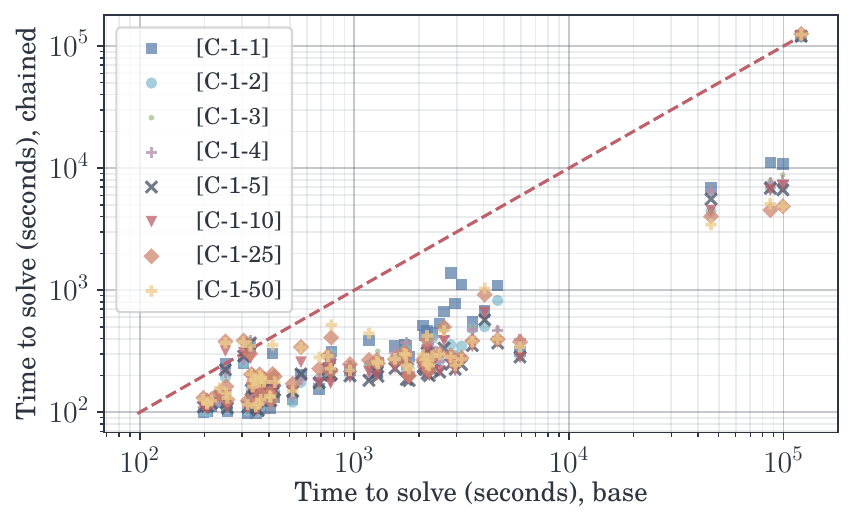}
    \caption{Comparison of time to solve PBES symbolically, including \pbeschain{} running time (log-log).}
    \label{fig:marijke-log-log-maxpvi}
  \end{subfigure}
  \caption{}
\end{figure}

For \casestudy{2} using the explicit solver~(\cref{fig:maeslant-maxpvi}), allowing up to 2 or 3 PVI in the unfolded formula does not speed up solving times compared to 1 PVI per unfolding.
In contrast, for \casestudy{1} in most cases, an increase in the number of allowed new PVI per formula speeds up solving the PBES.
The total sum to solve all chained properties is 198h for \casestudy{1-50} compared to 230h for \casestudy{1-1}.
This is not due to a lower number of PVI, as for \casestudy{1-50} there are on average 30\% more PVI after chaining, while for \casestudy{1-1} there are 40\% less PVI after chaining.
Instead, the number of nodes in the relevancy graph is consistently 7-10x lower when the maximum number of PVI increases. This has implications for memory use as well, which we discuss in the next section.
However, the time \pbeschain{} is running increases as the maximum number of PVI per unfolding increases. \Cref{fig:marijke-log-log-maxpvi} shows that for the PBESs that are faster to solve, the total time to solve (so including time to chain) can be higher than~\casestudy{1-1}, sometimes slower than not chaining at all and solving immediately.
In general, there is a tipping point when the benefit of chaining with a higher PVI bound is outweighed by overhead in time to chain.
For this model, the `left' region~($t < 10^3$) shows~\casestudy{1-1} lower (and therefore faster) than the other cases, the `middle' region~($10^3 \leq t \leq 10^4$) has~\casestudy{1-5} as the fastest chained PBESs, while \casestudy{1-50} comes out fastest for the `right' region~($t > 10^4$).

\paragraph{Standard Recursive Form (H3).}
This heuristic transforms a PBES to SRF before chaining when using the symbolic PBES solver.
For~\casestudy{2}, all the requirements are either just as fast or slightly slower to solve when applying SRF before chaining (i.e. 85s instead of 71s). We suspect that manipulating quantifiers when converting the PBES into SRF underlies minor additional overhead in the exploration phase of the symbolic solver. We can notice this here since this case study generally contains more complex data expressions.
For~\casestudy{1}, all requirements except one have the same time to solve and time to chain.
For~\casestudy{1}, requirement \lstinline|command_paddles_open| takes 10 minutes to solve after chaining when SRF transformation is performed before chaining, but 28 hours when no SRF transformation is performed before chaining.
The difference in the PBESs between both is that the no-SRF PBES contains a conjunct with a large data expression guarding it, while in the SRF PBES this expression is distributed over several PVIs. Large data expressions can cause additional computational overhead when the PBES gets explored when solving.
This case demonstrates that transforming a PBES to SRF beforehand can avoid computational overhead encountered when instead a PBES is transformed to SRF later.

\begin{figure}[hbtp]
  \centering
  \begin{subfigure}[t]{0.49\linewidth}
    \includegraphics[width=\textwidth]{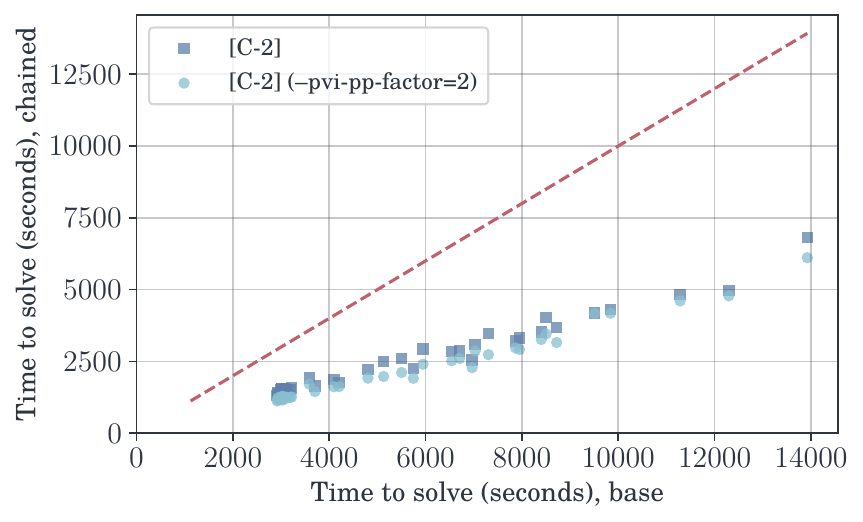}
    \caption{Only solving.}
  \end{subfigure}
  \begin{subfigure}[t]{0.49\linewidth}
    \includegraphics[width=\textwidth]{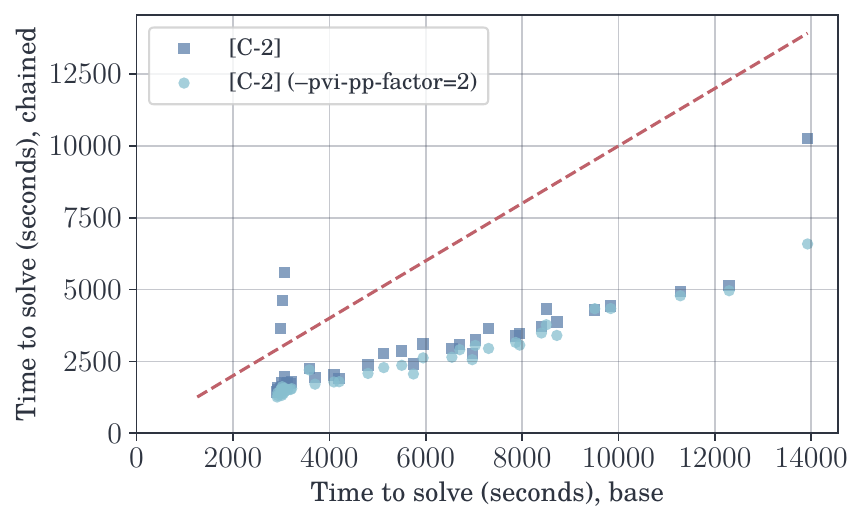}
    \caption{\pbeschain{} and solving.}
  \end{subfigure}
  \caption{Comparison of time to solve PBES before and after chaining.}
  \label{fig:maeslant-pvi-factor}
\end{figure}

\paragraph{Size of Unfolded Formula (H4).}
The goal of this heuristic is to avoid replacing PVIs with large formulae.
For~\casestudy{1} there is no difference when using factor 2: the difference in time to solve or time to chain when the heuristic is enabled is within the margin of noise.
In~\cref{fig:maeslant-pvi-factor}, we can see the effect this has for the symbolic solver on \casestudy{2}.
The heuristic significantly lowers the time to solve after chaining.
The most noticeable example in the benchmark data is requirement \lstinline|2_devices-3_ballast-ballast_valves_pumps| (the right-most dots in the figure), where the heuristic avoids unfolding to a large $\forall$-expression of roughly 2M characters. Without the bound, statically rewriting and analysing the chained PBES takes considerably longer. The general observation is that the pretty-print factor limits preprocessing overhead when an unfolding is syntactically large due to large data expressions, in this case present in the requirements for~\casestudy{2}.




\subsection{General trends}
We applied the heuristics to our case studies (Bound PVI Expansion = 5 for \casestudy{1} and 1 for \casestudy{2},\casestudy{3}, SRF for symbolic cases, Unfolded Formula Size Factor = 2 for \casestudy{2}).
Under these heuristics, there is a substantial reduction in the time it takes to check a property after chaining.
We also empirically confirm the validity of~\cref{theorem:relevancy-graph-srf}, since there is never an increase in relevant nodes after chaining.
For~\casestudy{2}, node counts before and after chaining exhibit a strong linear relationship, both for the explicit solver
and for the symbolic solver.
\casestudy{1}~displays a similar reduction pattern.
In terms of time to solve, for the symbolic solver, the largest reductions occur for requirements with the largest unchained solving times. For instance, requirement \lstinline|requirement21| first took 1h 40m to solve, but this was reduced to 4~minutes after chaining.
The biggest absolute speed-up occurred with \lstinline|command_entering_lights|, taking 1h 48m instead of 29 hours.
The properties that took the least time to solve without \pbeschain{} required around 200~seconds, while 24~chained PBESs took less than 100~seconds.
For \casestudy{2}, the time to check \lstinline|2_devices-1_dock-2_door-dock_catch| reduces 2.5~hours to 1~hour.
The explicit solver shows a very consistent pattern: each property for \casestudy{2} was 2.3-2.5 times faster to solve after chaining.
For \casestudy{3}, it took 1032.62 seconds to solve all properties without chaining and 930.79 seconds with chaining. This case study profits less from chaining before solving.

\begin{figure}[btp]
  \centering
  \begin{subfigure}[t]{0.49\linewidth}
    \includegraphics[width=\textwidth]{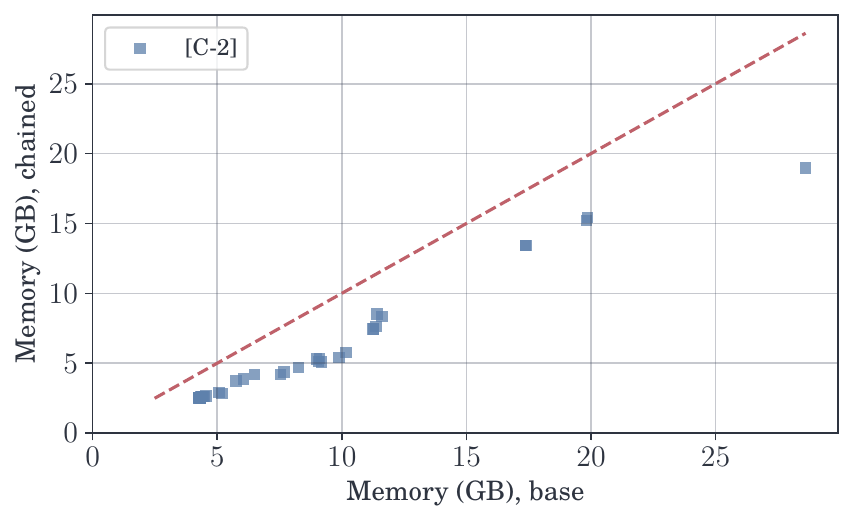}
    \caption{Explicit for \casestudy{2}.}
    \label{fig:maeslant-memory}
  \end{subfigure}
  \hfill
  \begin{subfigure}[t]{0.49\linewidth}
    \includegraphics[width=\textwidth]{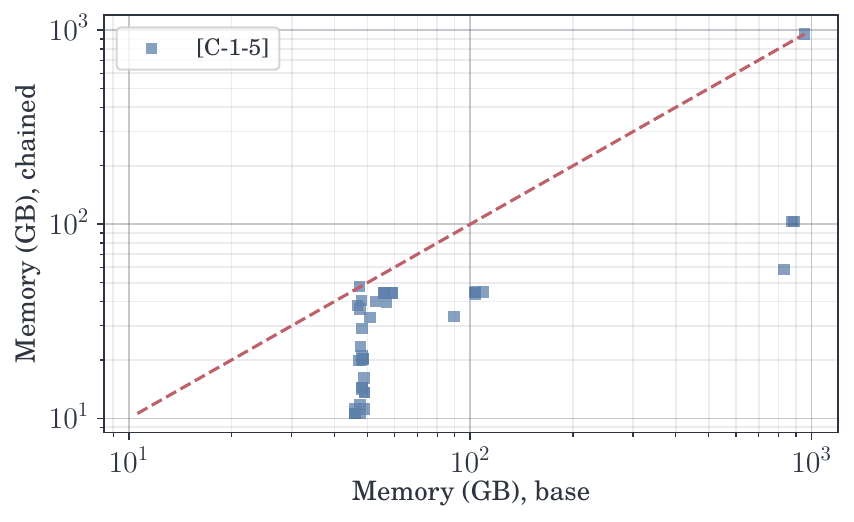}
    \caption{Symbolic for \casestudy{1-5} (log-log).}
    \label{fig:marijke-memory}
  \end{subfigure}
  \caption{Comparison of maximum memory used to solve PBES before and after chaining.}
  \label{fig:memory-computations}
\end{figure}

\paragraph{Memory computations.}
Besides computation time, also used memory is reduced with chaining.
We can compare the memory needed by a chained PBES for solving with the memory needed when solving the original PBES. We selected requirements covering the symbolic and explicit solver configurations and the available case studies. To this end, we set the memory limit argument of the symbolic solver to 1TB and looked at the maximum resident set size as provided by the GNU \tool{time} tool. The results can be seen in \cref{fig:memory-computations}. There is a clear trend that less memory is needed to solve chained PBESs, allowing these PBESs to be solved using less powerful systems.

\section{Concluding remarks}

When trying to establish correctness of critical software using model checking, the state space explosion problem can manifest itself. It is therefore desirable to have a plethora of methods that help mitigate this problem. We introduced one such method, called selective chaining, which works on PBESs, a framework for solving a variety of model checking problems.
We showed using real-world systems that selective chaining allows PBES solvers to reduce the time and memory space it needs to answer the encoded model checking problem. The benefit is up to 25x less time to solve and up to 14x less memory used. This enables solving model checking problems on machines with smaller memory capacity.
The provided heuristics help avoid cases where regular chaining operations can cause additional overhead.

Considering future work,
the order in which our algorithm chains PVIs currently does not follow a specific strategy.
The possibilities in terms of the order of PVI to chain remains unexplored, and since the order of chaining can change the final resulting PBES, it makes sense to explore static techniques that prioritize certain PVIs for chaining.

\printbibliography

\clearpage
\appendix
\section{Proofs for Section 3}

\subsection{Proofs for (\ref{sec:chaining_operator}) \nameref{sec:chaining_operator}}

\simultaneousSubstitutionEmpty*

\begin{proof}
  \setlength{\parindent}{0pt}
  We prove this using structural induction on $\varphi$.

  Consider $\varphi=b$. Then $b\isubsim{j}{\psi}{\varnothing}=b$ by definition.

  Consider $\varphi=Y(\vec e)$. Since $1\notin\varnothing$, the PVI is unchanged:
  $Y(\vec e)\isubsim{j}{\psi}{\varnothing}=Y(\vec e)$.

  Consider $\varphi=\forall d\sort D.\varphi'$; the case for $\varphi=\exists d\sort D.\varphi'$ is similar. Then:
  \[
    \begin{split}
      & (\forall d\sort D.\varphi')\isubsim{j}{\psi}{\varnothing} \\
      &= \forall d\sort D.\varphi'\isubsim{j}{\psi}{\varnothing} \\
      & \qquad \reason{Structural induction} \\
      &= \forall d\sort D.\varphi' \\
    \end{split}
  \]

  Consider $\varphi=\varphi_1\land\varphi_2$; the case for $\varphi=\varphi_1\lor\varphi_2$ is similar. Then:
  \[
    \begin{split}
      & (\varphi_1\land\varphi_2)\isubsim{j}{\psi}{\varnothing} \\
      &= \varphi_1\isubsim{j}{\psi}{\lc{i\in\varnothing\mid i\leq \npred(\varphi_1)}}
      \land
      \varphi_2\isubsim{j}{\psi}{\lc{i-\npred(\varphi_1)\mid i\in\varnothing\land i>\npred(\varphi_1)}} \\
      &= \varphi_1\isubsim{j}{\psi}{\varnothing}
      \land \varphi_2\isubsim{j}{\psi}{\varnothing} \\
      & \qquad \reason{Structural induction twice} \\
      &= \varphi_1\land\varphi_2 \\
    \end{split}
  \]
\end{proof}

\simultaneousMapstoidentity*

\begin{proof}
  \setlength{\parindent}{0pt}
  The case $I = \varnothing$ is covered in~\cref{lemma:simultaneous-substitution-empty}, so assume $I \neq \varnothing$.

  We prove this using structural induction on $\varphi$.
  The case $b$ holds trivially.
  For the cases $\forall d\sort D .\varphi$, and $\exists d\sort D .\varphi$, it is trivial using the induction hypothesis.

  Consider $Y(\vec e)\isub{i}{ \pv{Y(\vec e)}{i}}$.
  Then $i = 1$, $Y(\vec e) = \pv{Y(\vec e)}{i}$, and $I=\lc{1}$ so $Y(\vec e)\isubsim{i}{ \pv{Y(\vec e)}{i}}{I} = Y(\vec e)$.

  Let's consider the case $\varphi_1 \land \varphi_2$; the case for $\varphi_1 \lor \varphi_2$ is similar. We perform case distinction on~$i$. Consider when $\pviPositionsA{\pv{\varphi_1 \land \varphi_2}{i}}{\varphi_1 \land \varphi_2} \subseteq \npredset(\varphi_1)$, the case for $\pviPositionsA{\pv{\varphi_1 \land \varphi_2}{i}}{\varphi_1 \land \varphi_2} \cap \npredset(\varphi_1) = \varnothing$ is similar. Then $i \in \npredset(\varphi_1)$ and $I \subseteq \npredset(\varphi_1)$.
  \[
    \begin{split}
      & (\varphi_1\land\varphi_2)\isubsim{j}{\pv{\varphi_1 \land \varphi_2}{i}}{I} \\
      &= (\varphi_1\land\varphi_2)\isubsim{j}{\pv{\varphi_1}{i}}{I} \\
      &= \varphi_1\isubsim{j}{\pv{\varphi_1}{i}}{\lc{j\in I\mid j\leq \npred(\varphi_1)}} \\
      &\qquad \land
      \varphi_2\isubsim{j}{\pv{\varphi_1}{i}}{\lc{j-\npred(\varphi_1)\mid j\in I\land j>\npred(\varphi_1)}} \\
      & \qquad \reason{No overlap with $\varphi_2$} \\
      &= \varphi_1\isubsim{j}{\pv{\varphi_1}{i}}{\lc{j\in I\mid j\leq \npred(\varphi_1)}} \land
      \varphi_2 \\
      &= \varphi_1\isubsim{j}{\pv{\varphi_1}{i}}{I}
      \land \varphi_2\\
      & \qquad \reason{Structural induction} \\
      &= \varphi_1\land\varphi_2 \\
    \end{split}
  \]
  Consider when $\varnothing \subset \left(\pviPositionsA{\pv{\varphi_1 \land \varphi_2}{i}}{\varphi_1 \land \varphi_2} \cap \npredset(\varphi_1) \right) \subset \pviPositionsA{\pv{\varphi_1 \land \varphi_2}{i}}{\varphi_1 \land \varphi_2}$.
  Split $I$ into two disjoint sets $I' = I \cap \npredset(\varphi_1)$ and $I'' = I \setminus I'$.
  Assume $i \in I'$, the case is similar when  $i \in I''$.
  Then for all $i'' \in I''$ such that $i''\in \pviPositionsA{\pv{\varphi_1 \land \varphi_2}{i}}{\varphi_1 \land \varphi_2}$ we have by definition $\pv{\varphi_1 \land \varphi_2}{i} = \pv{\varphi_1 \land \varphi_2}{i''}$. Choose such an $i''$.
  \[
    \begin{split}
      & (\varphi_1\land\varphi_2)\isubsim{j}{\pv{\varphi_1 \land \varphi_2}{i}}{I} \\
      &= \varphi_1\isubsim{j}{\pv{\varphi_1 \land \varphi_2}{i}}{\lc{j\in I\mid j\leq \npred(\varphi_1)}} \\
      &\qquad \land
      \varphi_2\isubsim{j}{\pv{\varphi_1 \land \varphi_2}{i}}{\lc{j-\npred(\varphi_1)\mid j\in I\land j>\npred(\varphi_1)}} \\
      & \qquad \reason{Use $i'',I',I''$} \\
      &= \varphi_1\isubsim{j}{\pv{\varphi_1 \land \varphi_2}{i}}{I'} \land
      \varphi_2\isubsim{j}{\pv{\varphi_1 \land \varphi_2}{i''}}{\lc{j-\npred(\varphi_1)\mid j\in I''}} \\
      &= \varphi_1\isubsim{j}{\pv{\varphi_1}{i}}{I'} \land
      \varphi_2\isubsim{j}{\pv{\varphi_2}{i''-\npred(\varphi_1)}}{\lc{j-\npred(\varphi_1)\mid j\in I''}} \\
      & \qquad \reason{Structural induction twice} \\
      &= \varphi_1\land\varphi_2 \\
    \end{split}
  \]
\end{proof}

\subsection{Proofs for (\ref{sec:chaining_operator_correctness}) \nameref{sec:chaining_operator_correctness}}

\begin{proposition}
  \label{proposition:remove-unused-terms-from-delta}
  Let $\varphi$ be a predicate formula, and $\delta,\delta'$ be data environments such that for all $d\in \fv(\varphi)$, we have that $\delta(d) = \delta'(d)$.
  Then for all predicate environments $\eta$, we have that\\$\dbrack{\varphi}\eta\delta = \dbrack{\varphi}\eta\delta'$.

\end{proposition}

\begin{proof}
  \setlength{\parindent}{0pt}
  Choose some $\eta$. We perform induction on $\varphi$.

  Consider $\varphi = b$.
  By our definition of interpreting terms, we have for all free terms that the data environment evaluates to the same element. So this trivially holds.

  Consider $\varphi = {X(\vec e)}$. Then $\fv(X(\vec e)) = \fv(\vec e)$. Using the same reasoning as before,\[
    \dbrack{X(\vec e)}\eta\delta = \eta(X)(\dbrack{\vec e}\delta) \overs{}{=} \eta(X)(\dbrack{\vec e}\delta') = \dbrack{X(\vec e)}\eta\delta'
  \]

  Consider $\varphi = \forall d'\sort D'.\varphi'$; the case for $\exists d'\sort D'.\varphi'$ is similar. If $d' \notin \fv(\varphi')$, then the conclusion follows immediately from the induction hypothesis. So assume $d' \in \fv(\varphi')$.
  Then:
  \[
    \begin{split}
      &\dbrack{\forall d'\sort D'.\varphi'}\eta\delta \\
      & = \tforall{v'\in \mathbb{D}'} \dbrack{\varphi'}\eta\delta[v'/d'] \\
      & \qquad \reason{IH: for all $d \in \fv(\varphi')$ we know $\delta[v'/d'](d) = \delta'[v'/d'](d)$} \\
      & = \tforall{v'\in \mathbb{D}'} \dbrack{\varphi'}\eta\delta'[v'/d'] \\
      & = \dbrack{\forall d'\sort D'.\varphi'}\eta\delta'
    \end{split}
  \]

  Consider $\varphi = {\varphi_1 \land \varphi_2}$; the case for $\varphi_1 \lor \varphi_2$ is similar. Then:
  \[
    \begin{split}
      & \dbrack{\varphi_1 \land \varphi_2}\eta\delta \\
      & = \dbrack{\varphi_1}\eta\delta \tland \dbrack{\varphi_2}\eta\delta[v/d]\\
      & \qquad \reason{IH twice: $\fv(\varphi_1) \subseteq \fv(\varphi_1 \land \varphi_2)$ and $\fv(\varphi_2) \subseteq \fv(\varphi_1 \land \varphi_2)$}\\
      & = \dbrack{\varphi_1}\eta\delta' \tland \dbrack{\varphi_2}\eta\delta'[v/d]\\
      & = \dbrack{\varphi_1 \land \varphi_2}\eta\delta'[v/d]\\
    \end{split}
  \]
\end{proof}

The following lemma is present in the literature~\cite{stramagliaModelCheckingMachineControl2025}.

\begin{lemma}
  \label{lemma:from-delta-move-free-var-in-alt}
  Let $\varphi$ be a predicate formula, $d \in \fv(\varphi)$ be a data variable of sort~$D$
  and $e$~be a data term of sort~$D$.
  Then for all environments $\eta,\delta$, we have that $\dbrack{\varphi}\eta\delta
  [\dbrack{e}\delta/d]
  = \dbrack{\varphi\exprsub{d}{e}}\eta\delta$.
\end{lemma}

\begin{proof}
  Follows by induction on $\varphi$.
\end{proof}

From these lemmata, our \nameCref{lemma:from-delta-move-free-var-in} follows.

\fromdeltamovefreevarin*

\begin{proof}
  Since $\fv(\varphi) \subseteq \lc{\vec{d}}$,
  by \cref{proposition:remove-unused-terms-from-delta}, we can replace a data environment of the form $\delta'[\vec{v}/\vec{d}]$ with $\delta[\vec{v}/\vec{d}]$ when interpreting $\varphi$. The rest follows from \cref{lemma:from-delta-move-free-var-in-alt}.
\end{proof}

\begin{myomit}
\begin{remark}
  \label{remark:mapstointerpdelta}
  The following lemma fails:

  "Let $\varphi,\psi,\psi'$ be predicate formulae, $i \in \npredset(\varphi)$, $\delta$ be a data environment and $\eta,\eta'$ be predicate environments such that $\eta \subseteq \eta'$.
  Assume that $\dbrack{\psi}\eta\delta \timplies \dbrack{\psi'}\eta'\delta$.
  Then we have that $\dbrack{\varphi\isub{i}{ \psi}}\eta\delta \timplies \dbrack{\varphi\isub{i}{ \psi'}}\eta'\delta$."

  Let $\varphi \seteq \left(\forall n. (n = 0 \implies X(d))\right)$, $\psi \seteq (n=0)$ and $\psi' \seteq X(n)$. Let $i = 1$ and $\eta,\eta',\delta$ be such that $\eta=\eta'$, $\eta(X) = \lc{1}$, and $\delta(n) = 1$. Then:
  \[
    \begin{split}
      \dbrack{n=0}\eta\delta &= \false \\
      \dbrack{X(n)}\eta\delta &= \true \\
      \dbrack{\forall n. (n = 0 \implies n = 0)}\eta\delta &= \true \\
      \dbrack{\forall n. (n = 0 \implies X(n))}\eta\delta &= \false \\
    \end{split}
  \]
\end{remark}
\end{myomit}

\begin{myomit}
\mapstointerpdeltaimpwithpredicates*
\newcommand{\proofvd}{\vec{v_d}}
\newcommand{\proofd}{\vec{d}}

\begin{proof}
  \setlength{\parindent}{0pt}
  Pick some $\delta$ and $i \in \npredset(\varphi)$.
  We prove this using structural induction on $\varphi$.

  Consider $\varphi = b$. Then, $\dbrack{b\isub{i}{ \psi}}\eta\delta = \dbrack{b}\eta\delta$,  $\dbrack{b}\eta\delta \tequals \dbrack{b}\eta'\delta$, and $\dbrack{b}\eta'\delta = \dbrack{b\isub{i}{ \psi'}}\eta'\delta$.

  Consider $\varphi = Y(\vec e)$.
  Since $i=1$, then
  $\dbrack{Y(\vec e)\isub{i}{ \psi}}\eta\delta = \dbrack{\psi}\eta\delta$,
  $\dbrack{\psi}\eta\delta \timplies \dbrack{\psi'}\eta'\delta$ by assumption, and $\dbrack{\psi'}\eta'\delta = \dbrack{Y(\vec e)\isub{i}{ \psi'}}\eta'\delta$.

  Consider $\varphi = \forall d'\sort D'. \varphi'$; the case for $\varphi =\exists d'\sort D'. \varphi'$ is similar. For all $\delta'$ and $v_{d'} \in \mathbb{D}'$ we know that $\dbrack{\psi}\eta\delta'[v_{d'}/d'] \timplies \dbrack{\psi'}\eta'\delta'[v_{d'}/d']$ by assumption. Therefore:
  \[
    \begin{split}
      & \dbrack{(\forall d'\sort D'. \varphi')\isub{i}{ \psi}}\eta\delta \\
      &= \dbrack{\forall d'\sort D'. \varphi'\isub{i}{ \psi}}\eta\delta \\
      &= \tforall {v_{d'}\in\mathbb{D}'}\dbrack{\varphi'\isub{i}{ \psi}}\eta\delta[v_{d'}/d']  \\
      & \qquad \reason{IH} \\
      & \implies \tforall {v_{d'}\in\mathbb{D}'}\dbrack{\varphi'\isub{i}{ \psi'}}\eta'\delta[v_{d'}/d'] \\
      & = \dbrack{\forall d'\sort D'. \varphi'\isub{i}{ \psi'}}\eta'\delta \\
      &= \dbrack{(\forall d'\sort D'. \varphi')\isub{i}{ \psi'}}\eta'\delta
    \end{split}
  \]

  Consider $\varphi = \varphi_1 \land \varphi_2$; the case for $\varphi = \varphi_1 \lor \varphi_2$ is similar. If $i \in \npredset(\varphi_1)$, then:
  \[
    \begin{split}
      &\dbrack{(\varphi_1 \land \varphi_2)\isub{i}{ \psi}}\eta\delta \\
      &= \dbrack{\varphi_1\isub{i}{ \psi} \land \varphi_2}\eta\delta \\
      &= \dbrack{\varphi_1\isub{i}{ \psi}}\eta\delta \tland \dbrack{\varphi_2}\eta\delta  \\
      &\reason{By IH $\dbrack{\varphi_1\isub{i}{ \psi}}\eta\delta \implies\dbrack{\varphi_1\isub{i}{ \psi'}}\eta'\delta$, and monotonicity} \\
      &\implies \dbrack{\varphi_1\isub{i}{ \psi'}}\eta'\delta \tland \dbrack{\varphi_2}\eta'\delta \\
      &= \dbrack{\varphi_1\isub{i}{ \psi'} \land \varphi_2}\eta'\delta \\
      &= \dbrack{(\varphi_1 \land \varphi_2)\isub{i}{ \psi'}}\eta'\delta
    \end{split}
  \]

  If $i \notin \npredset(\varphi_1)$, then:
  \[
    \begin{split}
      &\dbrack{(\varphi_1 \land \varphi_2)\isub{i}{ \psi}}\eta\delta \\
      &= \dbrack{\varphi_1 \land \varphi_2\isub{(i - \npred(\varphi_1))}{ \psi}}\eta\delta \\
      &= \dbrack{\varphi_1}\eta\delta \tland \dbrack{\varphi_2\isub{(i - \npred(\varphi_1))}{ \psi}}\eta\delta  \\
      &\reason{Monotonicity and IH} \\
      &\implies \dbrack{\varphi_1}\eta'\delta \tland \dbrack{\varphi_2\isub{(i - \npred(\varphi_1))}{ \psi'}}\eta'\delta \\
      &= \dbrack{\varphi_1 \land \varphi_2\isub{(i - \npred(\varphi_1))}{ \psi'}}\eta'\delta \\
      &= \dbrack{(\varphi_1 \land \varphi_2)\isub{i}{ \psi'}}\eta'\delta
    \end{split}
  \]
\end{proof}

\end{myomit}

\begin{myomit}
\begin{proof}
  \setlength{\parindent}{0pt}
  Pick some $\delta \in \Delta$.
  We prove this using structural induction on $\varphi$.

  First, $\dbrack{b\isub{i}{ \psi}}\eta\delta = \dbrack{b}\eta\delta$,  $\dbrack{b}\eta\delta \timplies \dbrack{b}\eta'\delta$ by monotonicity, and $\dbrack{b}\eta'\delta = \dbrack{b\isub{i}{ \psi'}}\eta'\delta$.

  Consider $\dbrack{Y(\vec e)\isub{i}{ \psi}}\eta\delta$. If $i=1$, then
  $\dbrack{Y(\vec e)\isub{i}{ \psi}}\eta\delta = \dbrack{\psi}\eta\delta$,  $\dbrack{\psi}\eta\delta \timplies \dbrack{\psi'}\eta'\delta$ by assumption, and $\dbrack{\psi'}\eta'\delta = \dbrack{Y(\vec e)\isub{i}{ \psi'}}\eta'\delta$.
  If $i > 1$, then $\dbrack{Y(\vec e)\isub{i}{ \psi}}\eta\delta = \dbrack{Y(\vec e)}\eta\delta$, $\dbrack{Y(\vec e)}\eta\delta \timplies \dbrack{Y(\vec e)}\eta'\delta$ by monotonicity, and $\dbrack{Y(\vec e)}\eta'\delta = \dbrack{Y(\vec e)\isub{i}{ \psi'}}\eta'\delta$.

  Consider $\dbrack{(\forall d'\sort D. \varphi')\isub{i}{ \psi}}\eta\delta$; the case for $\dbrack{(\exists d'\sort D. \varphi')\isub{i}{ \psi}}\eta\delta$ is similar. For all $v \in \mathbb{D}$ we know that $\dbrack{\psi}\eta\delta[v/d'] \timplies \dbrack{\psi'}\eta\delta[v/d']$ since $\delta[v/d'] \in \Delta$, and $\dbrack{\psi'}\eta\delta[v/d'] \timplies \dbrack{\psi'}\eta'\delta[v/d']$ by monotonicity. Therefore:
  \[
    \begin{split}
      & \dbrack{(\forall d'\sort D. \varphi')\isub{i}{ \psi}}\eta\delta = \dbrack{\forall d'\sort D. \varphi'\isub{i}{ \psi}}\eta\delta \\
      &= \tforall {v\in\mathbb{D}}\dbrack{\varphi'\isub{i}{ \psi}}\eta\delta[v/d']  \\
      & \qquad \reason{IH: $d'\in \fv(\varphi')$ and $d'\notin \fv(\forall d'\sort D. \varphi')$ so $d'$ does not override some fixed value} \\
      & \implies \tforall {v\in\mathbb{D}}\dbrack{\varphi'\isub{i}{ \psi'}}\eta'\delta[v/d'] \\
      & = \dbrack{\forall d'\sort D. \varphi'\isub{i}{ \psi'}}\eta'\delta = \dbrack{(\forall d'\sort D. \varphi')\isub{i}{ \psi'}}\eta'\delta
    \end{split}
  \]

  Consider $\dbrack{(\varphi_1 \land \varphi_2)\isub{i}{ \psi}}\eta\delta$; the case for $\dbrack{(\varphi_1 \lor \varphi_2)\isub{i}{ \psi}}\eta\delta$ is similar. If $i \in \npredset(\varphi_1)$, then:
  \[
    \begin{split}
      &\dbrack{(\varphi_1 \land \varphi_2)\isub{i}{ \psi}}\eta\delta = \dbrack{\varphi_1\isub{i}{ \psi} \land \varphi_2}\eta\delta \\
      &= \dbrack{\varphi_1\isub{i}{ \psi}}\eta\delta \tland \dbrack{\varphi_2}\eta\delta & \\
      &\reason{IH, so $\dbrack{\varphi_1\isub{i}{ \psi}}\eta\delta \implies\dbrack{\varphi_1\isub{i}{ \psi'}}\eta\delta$, and monotonicity} \\
      &\implies \dbrack{\varphi_1\isub{i}{ \psi'}}\eta'\delta \tland \dbrack{\varphi_2}\eta'\delta \\
      &= \dbrack{\varphi_1\isub{i}{ \psi'} \land \varphi_2}\eta'\delta = \dbrack{(\varphi_1 \land \varphi_2)\isub{i}{ \psi'}}\eta'\delta & \\
    \end{split}
  \]

  If $i \notin \npredset(\varphi_1)$, then:
  \[
    \begin{split}
      &\dbrack{(\varphi_1 \land \varphi_2)\isub{i}{ \psi}}\eta\delta = \dbrack{\varphi_1 \land \varphi_2\isub{(i - \npred(\varphi_1))}{ \psi}}\eta\delta \\
      &= \dbrack{\varphi_1}\eta\delta \tland \dbrack{\varphi_2\isub{(i - \npred(\varphi_1))}{ \psi}}\eta\delta & \\
      &\reason{Monotonicity and IH} \\
      &\implies \dbrack{\varphi_1}\eta'\delta \tland \dbrack{\varphi_2\isub{(i - \npred(\varphi_1))}{ \psi'}}\eta'\delta \\
      &= \dbrack{\varphi_1 \land \varphi_2\isub{(i - \npred(\varphi_1))}{ \psi'}}\eta'\delta = \dbrack{(\varphi_1 \land \varphi_2)\isub{i}{ \psi'}}\eta'\delta  \\
    \end{split}
  \]
\end{proof}
\end{myomit}

\simultaneousSubstitutionMonotonicity*

\begin{proof}
  \setlength{\parindent}{0pt}
  Pick some $\delta$ and $I \subseteq \npredset(\varphi)$.
  We prove this using structural induction on $\varphi$.

  Consider $\varphi = b$. Then, $\dbrack{b\isubsim{j}{ \psi}{I}}\eta\delta = \dbrack{b}\eta\delta$,  $\dbrack{b}\eta\delta \tequals \dbrack{b}\eta'\delta$, and $\dbrack{b}\eta'\delta = \dbrack{b\isubsim{j}{ \psi'}{I}}\eta'\delta$.

  Consider $\varphi = Y(\vec e)$; its only position is $1$. If $1 \in I$, then
  $\dbrack{Y(\vec e)\isubsim{j}{ \psi}{I}}\eta\delta = \dbrack{\psi}\eta\delta$,
  $\dbrack{\psi}\eta\delta \timplies \dbrack{\psi'}\eta'\delta$ by assumption, and $\dbrack{\psi'}\eta'\delta = \dbrack{Y(\vec e)\isubsim{j}{ \psi'}{I}}\eta'\delta$.
  If $1 \notin I$, then $\dbrack{Y(\vec e)\isubsim{j}{ \psi}{I}}\eta\delta = \dbrack{Y(\vec e)}\eta\delta$, $\dbrack{Y(\vec e)}\eta\delta \timplies \dbrack{Y(\vec e)}\eta'\delta$ by monotonicity, and $\dbrack{Y(\vec e)}\eta'\delta = \dbrack{Y(\vec e)\isubsim{j}{ \psi'}{I}}\eta'\delta$.

  Consider $\varphi = \forall d'\sort D'. \varphi'$; the case for $\varphi =\exists d'\sort D'. \varphi'$ is similar. For all $\delta'$ and $v_{d'} \in \mathbb{D}'$ we know that $\dbrack{\psi}\eta\delta'[v_{d'}/d'] \timplies \dbrack{\psi'}\eta'\delta'[v_{d'}/d']$ by assumption. Therefore:
  \[
    \begin{split}
      & \dbrack{(\forall d'\sort D'. \varphi')\isubsim{j}{ \psi}{I}}\eta\delta \\
      &= \dbrack{\forall d'\sort D'. \varphi'\isubsim{j}{ \psi}{I}}\eta\delta \\
      &= \tforall {v_{d'}\in\mathbb{D}'}\dbrack{\varphi'\isubsim{j}{ \psi}{I}}\eta\delta[v_{d'}/d']  \\
      & \qquad \reason{IH} \\
      & \implies \tforall {v_{d'}\in\mathbb{D}'}\dbrack{\varphi'\isubsim{j}{ \psi'}{I}}\eta'\delta[v_{d'}/d'] \\
      & = \dbrack{\forall d'\sort D'. \varphi'\isubsim{j}{ \psi'}{I}}\eta'\delta \\
      &= \dbrack{(\forall d'\sort D'. \varphi')\isubsim{j}{ \psi'}{I}}\eta'\delta
    \end{split}
  \]

  Consider $\varphi = \varphi_1 \land \varphi_2$; the case for $\varphi_1 \lor \varphi_2$ is similar. Write $I_1=\lc{i\in I\mid i\leq \npred(\varphi_1)}$ and $I_2=\lc{i-\npred(\varphi_1)\mid i\in I\land i>\npred(\varphi_1)}$ for the position sets passed to the two operands. Then:
  \[
    \begin{split}
      &\dbrack{(\varphi_1 \land \varphi_2)\isubsim{j}{ \psi}{I}}\eta\delta \\
      &= \dbrack{\varphi_1\isubsim{j}{ \psi}{I_1} \land \varphi_2\isubsim{j}{ \psi}{I_2}}\eta\delta \\
      &= \dbrack{\varphi_1\isubsim{j}{ \psi}{I_1}}\eta\delta \tland \dbrack{\varphi_2\isubsim{j}{ \psi}{I_2}}\eta\delta  \\
      &\reason{By IH twice} \\
      &\implies \dbrack{\varphi_1\isubsim{j}{ \psi'}{I_1}}\eta'\delta \tland \dbrack{\varphi_2\isubsim{j}{ \psi'}{I_2}}\eta'\delta \\
      &= \dbrack{(\varphi_1 \land \varphi_2)\isubsim{j}{ \psi'}{I}}\eta'\delta
    \end{split}
  \]
\end{proof}

\newcommand{\etaneg}{\eta^\sqcup}
\newcommand{\etapos}{\eta^\sqcap}

\srfrelevantiff*

\begin{proof}
  We consider the disjunctive case, the conjunctive case is similar.

  \paragraph{Disjunctive}
  \emph{Case if}: Then there exists an $\eta,\delta$ s.t.
  \[
    \dbrack{\varphi_X}\eta\pvisub{\top}{Y(\vec{w})}\delta[\vec{v}/\vec{d}] \neq \dbrack{\varphi_X}\eta\pvisub{\bot}{Y(\vec{w})}\delta[\vec{v}/\vec{d}]
  \]
  So the former holds and the latter does not. Thus there exists a $j\in J_X, v_j \in \mathbb{E}_j$ such that $\dbrack{f_j(\vec{d},e_j)}\delta[\vec{v}/\vec{d}][v_j/e_j]$ holds and $\dbrack{X_j(\vec{g}_j(\vec{d},e_j))}\eta\pvisub{\top}{Y(\vec{w})}\delta[\vec{v}/\vec{d}][v_j/e_j]$ holds, while \[\dbrack{X_j(\vec{g}_j(\vec{d},e_j))}\eta\pvisub{\bot}{Y(\vec{w})}\delta[\vec{v}/\vec{d}][v_j/e_j]\] must not.
  Then it must be that $Y = X_j$ and $\vec{w}=\dbrack{\vec{g}_j(\vec{d},e_j)}\delta[\vec{v}/\vec{d}][v_j/e_j]$.

  \emph{Case only if}: Assume for some $j\in J_X, v_j \in \mathbb{E}_j$ we have $\dbrack{f_j(\vec{d},e_j)}\delta[\vec{v}/\vec{d}][v_j/e_j]$ holds, $Y = X_j$, and $\vec{w} = \dbrack{\vec{g}_j(\vec{d},e_j)}\delta[\vec{v}/\vec{d}][v_j/e_j]$. Choose $\etaneg$ such that $\etaneg(X')(v')$ never holds. Then $\dbrack{X_j(\vec{g}_j(\vec{d},e_j))}\etaneg\pvisub{\top}{Y(\vec{w})}\delta[\vec{v}/\vec{d}][v_j/e_j]$ holds, and:
  \[
    \begin{split}
      &\dbrack{X_j(\vec{g}_j(\vec{d},e_j))}\etaneg\pvisub{\top}{Y(\vec{w})}\delta[\vec{v}/\vec{d}][v_j/e_j] \text{ holds} \\
      &\timplies \\
      &\dbrack{f_j(\vec{d},e_j)}\delta[\vec{v}/\vec{d}][v_j/e_j] \tland \dbrack{X_j(\vec{g}_j(\vec{d},e_j))}\etaneg\pvisub{\top}{Y(\vec{w})}\delta[\vec{v}/\vec{d}][v_j/e_j]  \text{ holds} \\
      &\timplies \\
      & \dbrack{f_j(\vec{d},e_j) \land X_j(\vec{g}_j(\vec{d},e_j))}\etaneg\pvisub{\top}{Y(\vec{w})}\delta[\vec{v}/\vec{d}][v_j/e_j] \text{ holds} \\
      &\timplies \\
      & \dbrack{\bigvee_{j \in J_i} \exists e_j \sort E_j. f_j(\vec{d},e_j) \land X_j(\vec{g}_j(\vec{d},e_j))}\etaneg\pvisub{\top}{Y(\vec{w})}\delta[\vec{v}/\vec{d}] \text{ holds}
    \end{split}
  \]
  We can also deduce that $\dbrack{\varphi_X}\etaneg[\bot/Y(\vec{w})]\delta[\vec{v}/\vec{d}]$ does not hold.

  \paragraph{Conjunctive}
  \emph{Case if}: Then there exists an $\eta,\delta$ s.t.
  \[
    \dbrack{\varphi_X}\eta\pvisub{\top}{Y(\vec{w})}\delta[\vec{v}/\vec{d}] \neq \dbrack{\varphi_X}\eta\pvisub{\bot}{Y(\vec{w})}\delta[\vec{v}/\vec{d}]
  \]
  So the former holds and the latter does not. Thus there exists a $j\in J_X, v_j \in \mathbb{E}_j$ such that $\dbrack{f_j(\vec{d},e_j)}\delta[\vec{v}/\vec{d}][v_j/e_j]$ holds and $\dbrack{X_j(\vec{g}_j(\vec{d},e_j))}\eta\pvisub{\bot}{Y(\vec{w})}\delta[\vec{v}/\vec{d}][v_j/e_j]$ does not hold, while \[\dbrack{X_j(\vec{g}_j(\vec{d},e_j))}\eta\pvisub{\top}{Y(\vec{w})}\delta[\vec{v}/\vec{d}][v_j/e_j]\] holds.
  Then it must be that $Y = X_j$ and $\vec{w}=\dbrack{\vec{g}_j(\vec{d},e_j)}\delta[\vec{v}/\vec{d}][v_j/e_j]$.

  \emph{Case only if}: Then for some $j\in J_X, v_j \in \mathbb{E}_j$ we have $\dbrack{f_j(\vec{d},e_j)}\delta[\vec{v}/\vec{d}][v_j/e_j]$ holds, $Y = X_j$, and $\vec{w} = \dbrack{\vec{g}_j(\vec{d},e_j)}\delta[\vec{v}/\vec{d}][v_j/e_j]$. Choose $\etapos$ such that $\etapos(X')(v')$ always holds. Then $\dbrack{X_j(\vec{g}_j(\vec{d},e_j))}\etapos\pvisub{\bot}{Y(\vec{w})}\delta[\vec{v}/\vec{d}][v_j/e_j]$ does not hold, and:
  \[
    \begin{split}
      &\dbrack{X_j(\vec{g}_j(\vec{d},e_j))}\etapos\pvisub{\bot}{Y(\vec{w})}\delta[\vec{v}/\vec{d}][v_j/e_j] \text{ does not hold} \\
      &\timplies \\
      &\dbrack{f_j(\vec{d},e_j)}\delta[\vec{v}/\vec{d}][v_j/e_j] \timplies \dbrack{X_j(\vec{g}_j(\vec{d},e_j))}\etapos\pvisub{\bot}{Y(\vec{w})}\delta[\vec{v}/\vec{d}][v_j/e_j]  \text{ does not hold} \\
      &\timplies \\
      & \dbrack{f_j(\vec{d},e_j) \implies X_j(\vec{g}_j(\vec{d},e_j))}\etapos\pvisub{\bot}{Y(\vec{w})}\delta[\vec{v}/\vec{d}][v_j/e_j] \text{ does not hold} \\
      &\timplies \\
      & \dbrack{\bigwedge_{j \in J_i} \forall e_j \sort E_j. f_j(\vec{d},e_j) \implies X_j(\vec{g}_j(\vec{d},e_j))}\etapos\pvisub{\bot}{Y(\vec{w})}\delta[\vec{v}/\vec{d}] \text{ does not hold}
    \end{split}
  \]
  We can also deduce that $\dbrack{\varphi_X}\etapos[\top/Y(\vec{w})]\delta[\vec{v}/\vec{d}]$ holds.
\end{proof}

\ifapxrelevancygraphsrf
\relevancygraphsrf*
\proofrelevancygraphsrf
\fi

\ifapxpropproofgraph
\simultaneousChainingTrueDependencyGraph*
\proofpropproofgraph
\fi

\ifapxthmproofgraph
\simultaneousChainingProofGraph*
\proofthmproofgraph
\fi

\simnonewrelevanctedges*
\begin{proof}
  \setlength{\parindent}{0pt}
  We know that since $(Y,\vec{w}) \notin (X,\vec{v})^\bullet$ that for all $\eta$ that:
  \[
    \label{inner:relevant-phi}
    \tag{*}
    \dbrack{\varphi_X}\eta\pvisub{\top}{Y(\vec{w})}\delta[\vec{v}/\vec{d}] = \dbrack{\varphi_X}\eta\pvisub{\bot}{Y(\vec{w})}\delta[\vec{v}/\vec{d}]
  \]
  Also, we know that if for some $\delta$,
  $(X, \dbrack{\data{\varphi_X}{i}}\delta[\vec{v}/\vec{d}]) \in (X,\vec{v})^\bullet$
  then for all $\eta$ we have \[
    \label{inner:relevant-psi}
    \tag{**}
    \dbrack{\varphi_X\fvsub{\data{\varphi_X}{i}}}\eta\pvisub{\top}{Y(\vec{w})}\delta[\vec{v}/\vec{d}] = \dbrack{\varphi_X\fvsub{\data{\varphi_X}{i}}}\eta\pvisub{\bot}{Y(\vec{w})}\delta[\vec{v}/\vec{d}]
  \]
  We will show that then for all $\eta,\delta$ we have that \[
    \dbrack{\varphi_X\isubsim{i}{ \varphi_X\fvsub{\data{\varphi_X}{i}}}{I}}\eta\pvisub{\top}{Y(\vec{w})}\delta[\vec{v}/\vec{d}] = \dbrack{\varphi_X\isubsim{i}{ \varphi_X\fvsub{\data{\varphi_X}{i}}}{I}}\eta\pvisub{\bot}{Y(\vec{w})}\delta[\vec{v}/\vec{d}]
  \]
  \paragraph{Disjunctive}
  First consider the disjunctive SRF form, so:
  \[
    \varphi_X \seteq \bigvee_{j \in J_X} \exists e_j \sort E_j. f_j(\vec{d},e_j) \land X_j(\data{\varphi_X}{j})
  \]

  By contradiction. Assume there exists an $\eta,\delta$ such that
  \[
    \dbrack{\varphi_X\isubsim{i}{ \varphi_X\fvsub{\data{\varphi_X}{i}}}{I}}\eta\pvisub{\top}{Y(\vec{w})}\delta[\vec{v}/\vec{d}] \neq \dbrack{\varphi_X\isubsim{i}{ \varphi_X\fvsub{\data{\varphi_X}{i}}}{I}}\eta\pvisub{\bot}{Y(\vec{w})}\delta[\vec{v}/\vec{d}]
  \]
  Since all disjuncts except those in $I$ are the same in ${\varphi_X\isubsim{i}{ \varphi_X\fvsub{\data{\varphi_X}{i}}}{I}}$, by~(\ref{inner:relevant-phi}) it must be that there is some $j\in I$:
  \[
    \begin{split}
      & \dbrack{\exists e_j \sort E_j. f_j(\vec{d},e_j) \land \varphi_X\fvsub{\data{\varphi_X}{j}}}\eta\pvisub{\top}{Y(\vec{w})}\delta[\vec{v}/\vec{d}] \\
      &\neq \\
      & \dbrack{\exists e_j \sort E_j. f_j(\vec{d},e_j) \land \varphi_X\fvsub{\data{\varphi_X}{j}}}\eta\pvisub{\bot}{Y(\vec{w})}\delta[\vec{v}/\vec{d}] \\
    \end{split}
  \]
  Since by monotonicity the latter implies the former, there is some $v_j \in \mathbb{E}_j$ such that $\dbrack{f_j(\vec{d},e_j)}\delta[\vec{v}/\vec{d}][v_j/e_j]$ holds, and since $\data{\varphi_X}{i} = \data{\varphi_X}{j}$ we have two facts:
  \[
    \begin{split}
      & \dbrack{\varphi_X\fvsub{\data{\varphi_X}{i}}}
      \eta\pvisub{\top}{Y(\vec{w})}
      \delta[\vec{v}/\vec{d}][v_j/e_j]
      \neq
      \dbrack{\varphi_X\fvsub{\data{\varphi_X}{i}}}
      \eta\pvisub{\bot}{Y(\vec{w})}
      \delta[\vec{v}/\vec{d}][v_j/e_j]
    \end{split}
  \]
  and by~\cref{lemma:srf-relevant-iff} it must be that $(X,\dbrack{\data{\varphi_X}{j}} \delta[\vec{v}/\vec{d}][v_j/e_j]) \in (X,\vec{v})^\bullet$.
  But these two facts contradict~(\ref{inner:relevant-psi}).

  \paragraph{Conjunctive}
  Next consider the conjunctive SRF form, so:
  \[
    \varphi_X \seteq \bigwedge_{j \in J_X} \forall e_j \sort E_j. f_j(\vec{d},e_j) \implies X_j(\data{\varphi_X}{j})
  \]

  By contradiction. Assume there exists an $\eta,\delta$ such that
  \[
    \dbrack{\varphi_X\isubsim{i}{ \varphi_X\fvsub{\data{\varphi_X}{i}}}{I}}\eta\pvisub{\top}{Y(\vec{w})}\delta[\vec{v}/\vec{d}] \neq \dbrack{\varphi_X\isubsim{i}{ \varphi_X\fvsub{\data{\varphi_X}{i}}}{I}}\eta\pvisub{\bot}{Y(\vec{w})}\delta[\vec{v}/\vec{d}]
  \]
  Since all conjuncts except those in $I$ are the same in ${\varphi_X\isubsim{i}{ \varphi_X\fvsub{\data{\varphi_X}{i}}}{I}}$, by~(\ref{inner:relevant-phi}) it must be that there is some $j\in I$:
  \[
    \begin{split}
      & \dbrack{\forall e_j \sort E_j. f_j(\vec{d},e_j) \implies \varphi_X\fvsub{\data{\varphi_X}{j}}}\eta\pvisub{\top}{Y(\vec{w})}\delta[\vec{v}/\vec{d}] \\
      &\neq \\
      & \dbrack{\forall e_j \sort E_j. f_j(\vec{d},e_j) \implies \varphi_X\fvsub{\data{\varphi_X}{j}}}\eta\pvisub{\bot}{Y(\vec{w})}\delta[\vec{v}/\vec{d}] \\
    \end{split}
  \]
  Since by monotonicity the latter implies the former, there is some $v_j \in \mathbb{E}_j$ such that $\dbrack{f_j(\vec{d},e_j)}\delta[\vec{v}/\vec{d}][v_j/e_j]$ holds, and since $\data{\varphi_X}{i} = \data{\varphi_X}{j}$ we have two facts:
  \[
    \begin{split}
      & \dbrack{\varphi_X\fvsub{\data{\varphi_X}{i}}}
      \eta\pvisub{\top}{Y(\vec{w})}
      \delta[\vec{v}/\vec{d}][v_j/e_j]
      \neq
      \dbrack{\varphi_X\fvsub{\data{\varphi_X}{i}}}
      \eta\pvisub{\bot}{Y(\vec{w})}
      \delta[\vec{v}/\vec{d}][v_j/e_j]
    \end{split}
  \]
  and by~\cref{lemma:srf-relevant-iff} it must be that $(X,\dbrack{\data{\varphi_X}{j}} \delta[\vec{v}/\vec{d}][v_j/e_j]) \in (X,\vec{v})^\bullet$.
  But these two facts contradict~(\ref{inner:relevant-psi}).
\end{proof}

\ifapxsimrelevancygraphsrf
\simrelevancygraphsrf*

\proofsimrelevancygraphsrf
\fi

\clearpage

\section{Benchmark commands}
\label{appendix:benchmarks}

When running \lstinline|pbessolvesymbolic| during the benchmarks, all Maeslant Barrier had set the default memory limit of 3GB. For the Prinses Marijke lock complex, each property had been set a (higher) specific value needed for the instantiation step. These values can be found in~\cref{table:marijke-memory-limits}.

\subsection{Prinses Marijke lock complex}
\label{appendix:benchmark-marijke}
\begin{lstlisting}[language=bash,caption={The commands to verify the properties of the Prinses Marijke lock complex.}]
mcrl22lps {mcrl2path} {templpspath}
lpsfununfold {templpspath} | lpssuminst | lpsrewr | lpsconstelm | lpsparelm - {lpspath}
lps2pbes -f{propertypath} {lpspath} {temppbespath}
pbesconstelm -c {temppbespath} | pbesrewr | pbesparelm - | pbesrewr -pquantifier-inside | pbesrewr -pquantifier-one-point | pbesrewr | pbesrewr -pquantifier-all | pbesrewr | pbesconstelm -c | pbesrewr | pbesparelm | pbesconstelm -ce | pbesrewr - {pbespath}
pbeschain {pbespath} -rjittyp | pbesrewr - {pbeschainedpath}
pbessolvesymbolic {pbespath} --split-conditions --cached -rjittyc --groups=used --chaining --saturation --timings
pbessolvesymbolic {pbeschainedpath} --split-conditions --cached -rjittyc --groups=used --chaining --saturation --timings
\end{lstlisting}

\begin{table}[htbp]
  \footnotesize
  \centering
  \begin{tabular}{ll}
    \bfseries Property & \bfseries Memory limit \\\hline
    nodeadlock                  & 6   \\
    requirement1                & 8   \\
    requirement2                & 128 \\
    requirement3                & 64  \\
    requirement4                & 64  \\
    requirement5                & 8   \\
    requirement6                & 8   \\
    requirement7                & 64  \\
    requirement8                & 8   \\
    requirement9                & 8   \\
    requirement10               & 8   \\
    requirement11               & 32  \\
    requirement12               & 32  \\
    requirement13               & 8   \\
    requirement14               & 64  \\
    requirement15               & 64  \\
    requirement16               & 32  \\
    requirement17               & 64  \\
    requirement18               & 128 \\
    requirement19               & 8   \\
    requirement20               & 8   \\
    requirement23               & 64  \\
    requirement24               & 32  \\
    requirement27               & 8   \\
    requirement28               & 16  \\
    requirement29               & 8   \\
    requirement30               & 8   \\
    requirement31               & 8   \\
    requirement32               & 8   \\
    requirement33               & 8   \\
    requirement34               & 8   \\
    requirement35               & 8   \\
    requirement36               & 8   \\
    requirement40               & 8   \\
    command\_barrier\_close     & 8   \\
    command\_barrier\_open      & 8   \\
    command\_barrier\_stop      & 8   \\
    command\_barrier\_lights    & 8   \\
    command\_barrier\_emergency & 8   \\
    command\_gates\_close       & 8   \\
    command\_gates\_open        & 32  \\
    command\_gates\_stop        & 8   \\
    command\_paddles\_close     & 8   \\
    command\_paddles\_stop      & 8   \\
    command\_paddles\_open      & 8   \\
    command\_lock\_emergency    & 8   \\
    command\_leaving\_lights    & 129 \\
    command\_entering\_lights   & 129 \\
    barrier\_can\_be\_closed    & 8   \\
    gate\_can\_be\_closed       & 8   \\
    ship\_can\_pass             & 129 \\
    gate\_stop\_lights\_to\_red & 64  \\
    lock\_emergency\_to\_red    & 64  \\
    barrier\_emergency\_to\_red & 64
  \end{tabular}
  \caption{The memory limits for each property of the Prinses Marijke lock complex.}
  \label{table:marijke-memory-limits}
\end{table}

\subsection{Maeslant Barrier}
\label{appendix:benchmark-maeslant}
\begin{lstlisting}[language=bash,caption={The commands to verify the properties of the Maeslant Barrier.}]
mcrl22lps {mcrl2path} {templpspath} -n -Q0
lpsconstelm {templpspath} | lpssumelm | lpsrewr | lpsparelm | lpsstategraph - -Q0 {lpspath}
lps2pbes -f{propertypath} {lpspath} {temppbespath} -m
pbesconstelm {temppbespath} | pbesparelm | pbesrewr | pbesrewr -pquantifier-inside | pbesrewr -pquantifier-one-point | pbesrewr | pbesconstelm | pbesstategraph -Q0 -rjittyc | pbesconstelm - {pbespath}
pbeschain {pbespath} -m3 --quantifier-free --pvi-pp-factor=2 | pbesrewr - {pbeschainedpath}
pbessolvesymbolic {pbespath} -Q0 -rjittyc --split-conditions --cached --groups=used
pbessolvesymbolic {pbeschainedpath} -Q0 -rjittyc --split-conditions --cached --groups=used
\end{lstlisting}

\subsection{Pneumatic Cylinder}
\label{appendix:benchmark-cylinder}
\begin{lstlisting}[language=bash,caption={The commands to verify the properties of the Pneumatic Cylinder.}]
mcrl22lps {mcrl2path} {lpspath} -nf
lps2pbes -f{propertypath} {lpspath} {pbespath}
pbeschain {pbespath} --quantifier-free --pvi-pp-factor=2 | pbesrewr - {pbeschainedpath}
pbessolve {pbespath} -Q0 -rjittyc 
pbessolve {pbeschainedpath} -Q0 -rjittyc
\end{lstlisting}

\end{document}